\documentclass[aps,prd,twocolumn,nofootinbib,showpacs,longbibliography]{revtex4-1}
\usepackage{bm,amsmath,amssymb,graphicx}
\usepackage[colorlinks,linkcolor=blue,urlcolor=blue]{hyperref}
\begin{document}

\title{Coordinate conditions and field equations for pure composite gravity}

\author{Hans Christian \"Ottinger}
\email[]{hco@mat.ethz.ch}
\homepage[]{www.polyphys.mat.ethz.ch}
\affiliation{ETH Z\"urich, Department of Materials, CH-8093 Z\"urich, Switzerland}

\date{\today}

\begin{abstract}
Whenever an alternative theory of gravity is formulated in a background Minkowski space, the conditions characterizing admissible coordinate systems, in which the alternative theory of gravity may be applied, play an important role. We here propose Lorentz covariant coordinate conditions for the composite theory of pure gravity developed from the Yang-Mills theory based on the Lorentz group, thereby completing this previously proposed higher derivative theory of gravity. The physically relevant static isotropic solutions are determined by various methods, the high-precision predictions of general relativity are reproduced, and an exact black-hole solution with mildly singular behavior is found.
\end{abstract}


\maketitle

\section{Introduction}
Only two years after the discovery of Yang-Mills theories \cite{YangMills54}, it has been recognized that that there is a striking formal relationship between the Riemann curvature tensor of general relativity and the field tensor of the Yang-Mills theory based on the Lorentz group \cite{Utiyama56}. However, developing this particular Yang-Mills theory into a consistent and convincing theory of gravity is not at all straightforward. The ideas of \cite{Utiyama56} have been found to be ``unnatural'' by Yang (see footnote 5 of \cite{Yang74}), whose work has later been criticized massively in Chap.~19 of \cite{BlagojevicHehl}. Nevertheless, the pioneering work \cite{Utiyama56} may be considered as the origin of what is now known as gauge gravitation theory \cite{CapozzielloDeLau11,IvanenkoSar83}.

An obvious problem with the Yang-Mills theory based on the Lorentz group is that it has the large number of $48$ degrees of freedom, half of which are physically relevant. One is faced with six four-vector fields satisfying second-order evolution equations. For the pure field theories, the physical degrees of freedom are essentially given by the two transverse components of the four-vector fields, like in electrodynamics with its single vector potential. In view of this enormous number of degrees of freedom we need an almost equally large number of constraints to keep only a few degrees of freedom in a theory of gravity. In other words, we need a structured principle for selecting just a few ones among all the solutions of the Yang-Mills theory based on the Lorentz group.

A powerful selection principle can be implemented by means of the tool of composite theories \cite{hco235,hco237}. The basic idea is to write the gauge vector fields of the Yang-Mills theory in terms of fewer, more fundamental variables and their derivatives. The admission of derivatives in this so-called composition rule implies that the composite theory involves higher than second derivatives. The power of the tool of composite theories results from the fact that, in their Hamiltonian formulations \cite{hco235,hco237}, the structure of the constraints providing the selection principle is highly transparent.

As the composite theory of gravity \cite{hco231}, just like the underlying Yang-Mills theory, is formulated in a background Minkowski space, the question arises how to characterize the ``good'' coordinate systems in which the theory may be applied. This characterization should be Lorentz invariant, but not invariant under more general coordinate transformations, that is, it shares the formal properties of coordinate conditions in general relativity. However, the unique solutions obtained from Einstein's field equations only after specifying coordinate conditions are all physically equivalent, whereas the coordinate conditions in composite gravity characterize physically preferred systems. From a historical perspective, it is remarkable that Einstein in 1914 still believed that the metric should be completely determined by the field equations and, therefore, a generally covariant theory of gravity was not desirable (see \cite{Giovanelli20} for a detailed discussion). The important task of characterizing the preferred systems in composite gravity is addressed in the present paper. Once it is solved, we can provide a canonical Hamiltonian formulation of composite theory of gravity beyond the weak-field approximation \cite{hco240} and we obtain the static isotropic black-hole solution in a proper coordinate system.

The structure of the paper is as follows. As a preparatory step, we present the various variables and relations between them (Sec.~\ref{secvarvar}) and discuss their gauge transformation behavior (Sec.~\ref{secgaugetrans}). A cornerstone of the development is the close relationship between the covariant derivatives associated (i) with a connection with torsion and (ii) with the Yang-Mills theory based on the Lorentz group. The core of the composite theory of gravity consists of the field equations presented for several sets of variables (Sec.~\ref{secfieldeqs}) and the coordinate conditions characterizing the admissible coordinate systems (Sec.~\ref{seccoco3}). As an application, we determine the static isotropic solutions and provide the results for the high-precision tests of gravity as well as an exact black-hole solution (Sec.~\ref{secisosol}). We finally offer a detailed summary of our results and draw a number of conclusions (Sec.~\ref{secconclusions}). A number of detailed results and arguments are provided in six appendices.

\section{Various variables and relations between them}\label{secvarvar}
For the understanding of composite theories, it is important to introduce different kinds of variables and to clarify the relations between them. On the one hand, we have the metric tensors, tetrad variables, connections and curvature tensors familiar from general relativity and other theories of gravity. On the other hand, we have the gauge vectors and field tensors of the Yang-Mills theory based on the Lorentz group.

An important step is the decomposition of metric tensors in terms of tetrad or \emph{vierbein} variables,
\begin{equation}\label{localinertialdecomp}
   g_{\mu\nu} = \eta_{\kappa\lambda} \, {b^\kappa}_\mu {b^\lambda}_\nu
   = {b^\kappa}_\mu \, b_{\kappa\nu} ,
\end{equation}
where $\eta_{\kappa\lambda} = \eta^{\kappa\lambda}$ is the Minkowski metric with signature $(-,+,+,+)$. Throughout this paper, the Minkowski metric is used for raising or lowering space-time indices. For the inverses of the metric and the tetrad variables we introduce the components $\bar{g}^{\mu\nu}$ and $\mbox{$\bar{b}^\mu$}_{\kappa}$. Note that they are not obtained by raising or lowering indices of $g_{\mu\nu}$ and ${b^\kappa}_\mu$, respectively.

Equation~(\ref{localinertialdecomp}) may be regarded as the characterization of metric tensors by symmetry and definiteness properties. A general metric tensor may also be regarded as the result of transforming the Minkowski metric. The decomposition of a metric $g_{\mu\nu}$ into tetrad variables ${b^\kappa}_\mu$ is not unique. If we multiply ${b^\kappa}_\mu$ from the left with any Lorentz transformation, the invariance of the Minkowski metric under Lorentz transformations implies that we obtain another valid decomposition. This observation reveals the origin of the underlying gauge symmetry of the composite theory of gravity.

The key role of the metric tensor in the present theory is the characterization of the momentum-velocity relation, so that it can be interpreted as an indication of tensorial properties of mass. While this is also the case in general relativity, Einstein's theory of gravity goes much further in the geometric interpretation of the metric by assuming that it characterizes the underlying space-time. In contrast, the present theory is developed in an underlying Minkowski space, which is the standard situation for Yang-Mills theories.

As a next step, we introduce the vector fields $A_{(\kappa\lambda) \rho}$ in terms of the tetrad variables (the pair $(\kappa, \lambda)$ of space-time indices should be considered as a label associated with the Lorentz group, $\rho$ as a four-vector index),
\begin{eqnarray}
   {b^\kappa}_\mu {b^\lambda}_\nu \, A_{(\kappa\lambda) \rho} &=&
   \frac{1}{2} \left( \frac{\partial g_{\nu\rho}}{\partial x^\mu}
   -\frac{\partial g_{\mu\rho}}{\partial x^\nu} \right)
   \nonumber\\
   &+& \frac{1}{2 \tilde{g}} 
   \left( {b^\kappa}_\mu \, \frac{\partial b_{\kappa\nu}}{\partial x^\rho}
   - \frac{\partial{b^\kappa}_\mu}{\partial x^\rho} \, b_{\kappa\nu}
   \right) . \qquad 
\label{nonlinearcompositioncu}
\end{eqnarray}
From the Yang-Mills perspective, $\tilde{g}$ is the coupling constant. From a metric viewpoint, $\tilde{g} \ne 1$ implies torsion (see Eq.~(\ref{Gammatildef}) below). The antisymmetry of the right-hand side of Eq.~(\ref{nonlinearcompositioncu}) in $\mu$ and $\nu$ leads, after resolving for $A_{(\kappa\lambda) \rho}$, to antisymmetry in $\kappa$ and $\lambda$. We have thus introduced six vector fields associated with six pairs $(\kappa,\lambda)$, or with a label $a$ taking the values from $1$ to $6$ according to Table~\ref{tabindexmatch}. The pairs $(0,1)$, $(0,2)$, $(0,3)$ correspond to Lorentz boosts in the respective directions (involving also time) and the pairs $(2,3)$, $(3,1)$, $(1,2)$ correspond to rotations in the respective planes, as can be recognized by analyzing the gauge transformation behavior of the fields $A_{(\kappa\lambda) \rho}$ resulting from the freedom of acting with Lorentz transformations on ${b^\kappa}_\mu$ (see Sec.~\ref{secgaugetrans} for details).

\begin{table}
\begin{tabular}{c|c c c c c c}
    $a$ \, & \, $1$ & $2$ & $3$ & $4$ & $5$ & $6$ \\
	\hline
	$(\kappa,\lambda)$ \,
    & \, $(0,1)$ & $(0,2)$ & $(0,3)$ & $(2,3)$ & $(3,1)$ & $(1,2)$ \\
\end{tabular}
\caption{Correspondence between the label $a$ for the base vectors of the six-dimensional Lie algebra ${\rm so}(1,3)$ and ordered pairs $(\kappa,\lambda)$ of space-time indices.}
\label{tabindexmatch}
\end{table}

Following standard procedures for Yang-Mills theories (see, e.g., Sect.~15.2 of \cite{PeskinSchroeder}, Chap.~15 of \cite{WeinbergQFT2}, or \cite{hco229}), we can introduce a field tensor in terms of the vector fields,
\begin{equation}\label{Fdefinition}
   F_{a \mu\nu} = \frac{\partial A_{a \nu}}{\partial x^\mu}
   - \frac{\partial A_{a \mu}}{\partial x^\nu}
   + \tilde{g} f^{bc}_a A_{b \mu} A_{c \nu} ,
\end{equation}
where $f^{bc}_a$ are the structure constants of the Lorentz group. A Lie algebra label, say $a$, can be raised or lowered by raising or lowering the indices in the pairs associated with $a$ according to Table~\ref{tabindexmatch}. The structure constants can then be specified as follows: $f^{abc}$ is $1$ ($-1$) if $(a,b,c)$ is an even (odd) permutation of $(4,5,6)$, $(1,3,5)$, $(1,6,2)$ or $(2,4,3)$ and $0$ otherwise (see also Eq.~(\ref{Lorentzstructure})).

The definition (\ref{nonlinearcompositioncu}) suggests the following general passage from quantities labeled by a Lie algebra index to a quantity with space-time indices,
\begin{equation}\label{Xtildef}
   \tilde{X}_{\mu\nu} = {b^\kappa}_\mu {b^\lambda}_\nu \, X_{(\kappa\lambda)} .
\end{equation}
One then gets a deep relation between covariant derivatives associated with metrics and connections on the one hand and covariant derivatives associated with a Yang-Mills theory based on the Lorentz group on the other hand (for a proof of this fundamental relation based on the structure of the Lorentz group, see Appendix~\ref{Appcovdevs}),
\begin{eqnarray}
   \frac{\partial \tilde{X}_{\mu\nu}}{\partial x^\rho}
   - \Gamma^\sigma_{\rho\mu} \tilde{X}_{\sigma\nu}
   - \Gamma^\sigma_{\rho\nu} \tilde{X}_{\mu\sigma} &=& \nonumber\\
   && \hspace{-9em} {b^\kappa}_\mu {b^\lambda}_\nu \left[
   \frac{\partial X_{(\kappa\lambda)}}{\partial x^\rho}
   + \tilde{g} \, f_{(\kappa\lambda)}^{bc} A_{b\rho} X_c \right] ,
\label{central}
\end{eqnarray}
where the connection $\Gamma^\rho_{\mu\nu}$ is given by
\begin{equation}\label{Gammatildef}
   \Gamma^\rho_{\mu\nu} = \frac{1}{2} \, \bar{g}^{\rho\sigma}
   \left[ \frac{\partial g_{\sigma\nu}}{\partial x^\mu} + \tilde{g}
   \left( \frac{\partial g_{\mu\sigma}}{\partial x^\nu}
   - \frac{\partial g_{\mu\nu}}{\partial x^\sigma} \right) \right] =
   \bar{g}^{\rho\sigma} \, \bar{\Gamma}_{\sigma\mu\nu} .
\end{equation}
Unlike the Christoffel symbols obtained for $\tilde{g} = 1$, $\Gamma^\rho_{\mu\nu}$ is not symmetric in $\mu$ and $\nu$ for $\tilde{g} \ne 1$. This lack of symmetry indicates the presence of torsion. Note, however, that the connection is metric-compatible for all $\tilde{g}$ \cite{Jimenezetal19}, that is,
\begin{equation}\label{metriccompatible}
   \frac{\partial g_{\mu\nu}}{\partial x^\rho}
   - \Gamma^\sigma_{\rho\mu} g_{\sigma\nu}
   - \Gamma^\sigma_{\rho\nu} g_{\mu\sigma} = 0 ,
\end{equation}
which can be recast in the convenient form
\begin{equation}\label{metriccompatiblex}
   \frac{\partial g_{\mu\nu}}{\partial x^\rho} =
   \bar{\Gamma}_{\mu\rho\nu} + \bar{\Gamma}_{\nu\rho\mu} .
\end{equation}
From the connection $\Gamma^\rho_{\mu\nu}$, we can further construct the Riemann curvature tensor (see, e.g. \cite{Jimenezetal19} or \cite{Weinberg})
\begin{equation}\label{R4def}
   {R^\mu}_{\nu\mu'\nu'} = \frac{\partial \Gamma^\mu_{\mu'\nu}}{\partial x^{\nu'}}
   - \frac{\partial \Gamma^\mu_{\nu'\nu}}{\partial x^{\mu'}}
   + \Gamma^\sigma_{\mu'\nu} \Gamma^\mu_{\nu'\sigma}
   - \Gamma^\sigma_{\nu'\nu} \Gamma^\mu_{\mu'\sigma} .
\end{equation}

In Appendix~\ref{Appfieldtensor}, it is shown that the field tensor (\ref{Fdefinition}) can be written in the alternative form
\begin{eqnarray}
   \tilde{F}_{\mu\nu \mu'\nu'} &=& \nonumber\\
   && \hspace{-2em} 
   \frac{1}{2} \left( 
   \frac{\partial^2 g_{\nu\nu'}}{\partial x^\mu \partial x^{\mu'}}
   - \frac{\partial^2 g_{\nu\mu'}}{\partial x^\mu \partial x^{\nu'}}
   - \frac{\partial^2 g_{\mu\nu'}}{\partial x^\nu \partial x^{\mu'}}
   + \frac{\partial^2 g_{\mu\mu'}}{\partial x^\nu \partial x^{\nu'}} \right)
   \nonumber\\
   && \hspace{-2em} + \, \frac{1}{\tilde{g}} \, \bar{g}^{\rho\sigma}
   ( \bar{\Gamma}_{\rho\mu'\mu} \bar{\Gamma}_{\sigma\nu'\nu}
   - \bar{\Gamma}_{\rho\nu'\mu} \bar{\Gamma}_{\sigma\mu'\nu} ) .
\label{Fdefinitiontilz}
\end{eqnarray}
This explicit expression for $\tilde{F}_{\mu\nu \mu'\nu'}$ reveals its symmetry properties: antisymmetry under $\mu \leftrightarrow \nu$ and $\mu' \leftrightarrow \nu'$ and, more surprisingly, symmetry under $(\mu\nu) \leftrightarrow (\mu'\nu')$. A comparison between the expressions (\ref{R4def}) and (\ref{Fdefinitiontilz}) yields a remarkable relationship between the Riemann curvature tensor and the field tensor of the Yang-Mills theory based on the Lorentz group,
\begin{equation}\label{Fdefinitiontily}
   \tilde{g} \, \bar{g}^{\mu\rho} \tilde{F}_{\rho\nu \mu'\nu'} =
   {R^\mu}_{\nu\mu'\nu'} ,
\end{equation}
which holds for all values of the coupling constant $\tilde{g}$.

\section{Gauge transformation behavior}\label{secgaugetrans}
As a consequence of the decomposition (\ref{localinertialdecomp}), there exists the gauge freedom of acting with a Lorentz transformation from the left on the tetrad variables ${b^\kappa}_\mu$. In its infinitesimal version, this possibility corresponds to the transformation
\begin{equation}\label{gaugeb}
   \delta b_{\kappa\mu} = \tilde{g} \, \Lambda_{(\kappa\lambda)} {b^\lambda}_\mu ,
\end{equation}
where $\Lambda_{(\kappa\lambda)}$ is antisymmetric in $\kappa$ and $\lambda$ and can hence be understood as $\Lambda_a$ according to Table~\ref{tabindexmatch}. For $\kappa=0$, time is mixed with a spatial dependence in one of the coordinate directions so that we deal with the respective Lorentz boosts. If both $\kappa=k$ and $\lambda=l$ are both spatial indices, the antisymmetric matrix $\Lambda_{(\kappa\lambda)}$ describes rotations in the corresponding $(k,l)$ plane. For the inverse of ${b^\kappa}_\mu$, Eq.~(\ref{gaugeb}) implies
\begin{equation}\label{gaugebbar}
   \delta\mbox{$\bar{b}^\mu$}_{\kappa} =
   - \tilde{g} \, \Lambda_{(\kappa\lambda)} \bar{b}^{\mu\lambda} .
\end{equation}

By using Eq.~(\ref{gaugeb}) in the composition rule (\ref{nonlinearcompositioncu}), we obtain
\begin{equation}\label{gaugeAp}
   \delta A_{(\kappa\lambda) \rho} - \tilde{g} \, \eta^{\kappa'\lambda'}
   \Big[ A_{(\kappa'\lambda) \rho} \Lambda_{(\kappa\lambda')}
   - \Lambda_{(\kappa'\lambda)} A_{(\kappa\lambda') \rho} \Big]
   = \frac{\partial \Lambda_{(\kappa\lambda)}}{\partial x^\rho} ,
\end{equation}
which, by means of Eq.~(\ref{supauxf1}), can be written as
\begin{equation}\label{gaugeA}
   \delta A_{a \rho} = \frac{\partial \Lambda_a}{\partial x^\rho}
   + \tilde{g} f^{bc}_a \, A_{b \rho} \, \Lambda_c .
\end{equation}
This result demonstrates that the six vector fields $A_{a \rho}$ indeed possess the proper gauge transformation behavior for the vector fields of the Yang-Mills theory based on the Lorentz group. By means of the Jacobi identity for the structure constants,
\begin{equation}\label{Jacobiid}
   f^{sb}_a f^{cd}_s + f^{sc}_a f^{db}_s + f^{sd}_a f^{bc}_s = 0 ,
\end{equation}
we further obtain the gauge transformation behavior of the field tensor,
\begin{equation}\label{gaugeF}
   \delta F_{a \mu\nu} = \tilde{g} f^{bc}_a \, F_{b \mu\nu} \, \Lambda_c .
\end{equation}

Finally, we look at the gauge transformation behavior obtained for the Yang-Mills variables transformed according to Eq.~(\ref{Xtildef}). From Eqs.~(\ref{gaugeb}) and (\ref{gaugeAp}) we obtain
\begin{equation}\label{gaugeAtil}
   \delta \tilde{A}_{\mu\nu\rho} = {b^\kappa}_\mu {b^\lambda}_\nu \, 
   \frac{\partial \Lambda_{(\kappa\lambda)}}{\partial x^\rho} .
\end{equation}
As the metric is gauge invariant (gauge degrees of freedom result only from its decomposition), the representations (\ref{Gammatildef}) and (\ref{Fdefinitiontilz}) imply the gauge invariance properties
\begin{equation}\label{gaugeGam}
   \delta \Gamma^\rho_{\mu\nu} = \delta \bar{\Gamma}_{\sigma\mu\nu} = 0 ,
\end{equation}
and
\begin{equation}\label{gaugeFtil}
   \delta \tilde{F}_{\mu\nu \mu'\nu'} = 0 .
\end{equation}

\section{Field equations}\label{secfieldeqs}
With the help of Eq.~(\ref{central}), the standard field equations for our Yang-Mills theory based on the Lorentz group (see, e.g., Sect.~15.2 of \cite{PeskinSchroeder}, Chap.~15 of \cite{WeinbergQFT2}, or \cite{hco229}) can be written in the manifestly gauge invariant form
\begin{equation}\label{YMfieldeqs}
   \eta^{\mu'\mu''} \left( \frac{\partial \tilde{F}_{\mu\nu \mu''\nu'}}{\partial x^{\mu'}}
   - \Gamma^\sigma_{\mu'\mu} \tilde{F}_{\sigma\nu \mu''\nu'}
   - \Gamma^\sigma_{\mu'\nu} \tilde{F}_{\mu\sigma \mu''\nu'} \right) = 0.
\end{equation}
By means of Eq.~(\ref{Fdefinitiontily}), these field equations can be rewritten in terms of the Riemann curvature tensor,
\begin{equation}\label{YMfieldeqsR}
   \eta^{\rho\nu'} \left( \frac{\partial {R^\mu}_{\nu\mu'\nu'}}{\partial x^\rho}
   + \Gamma^\mu_{\rho\sigma} {R^\sigma}_{\nu\mu'\nu'}
   - \Gamma^\sigma_{\rho\nu} {R^\mu}_{\sigma\mu'\nu'} \right) = 0.
\end{equation}
In view of Eq.~(\ref{R4def}), this latter equation is entirely in terms of the variables $\Gamma^\rho_{\mu\nu}$. The explicit form of the resulting equation is given in Appendix~\ref{Appfieldeqcon}. This observation offers the option of the following two-step procedure: one first determines the most general solution of the second-order differential equations (\ref{compactGameq}) for $\Gamma^\rho_{\mu\nu}$ and then, in a post-processing step, one obtains the metric by solving the first-order differential equations (\ref{Gammatildef}). The post-processing step selects those solutions $\Gamma^\rho_{\mu\nu}$ that can actually be expressed in terms of the metric. 

Finally, we write the field equations directly as third-order differential equations for the metric. As the solutions of these third-order equations can be understood in terms of selected solutions of the Yang-Mills theory found by post-processing, there is no reason to be concerned about the potential instabilities resulting from higher-order differential equations, known as Ostrogradsky instabilities \cite{Ostrogradsky1850,Woodard15}. Avoiding such instabilities is an important topic, in particular, in alternative theories of gravity \cite{Chenetal13,RaidalVeermae17,Stelle77,Stelle78,Krasnikov87,GrosseKnetter94,Beckeretal17,Salvio19}. We write all the third and second derivatives of the metric explicitly, whereas the first derivatives are conveniently combined into connection variables. The result is the following set of equations for the composite theory of gravity obtained by expressing the gauge vector fields of the Yang-Mills theory based on the Lorentz group in terms of the tetrad variables obtained by decomposing a metric,
\begin{eqnarray}
   \Xi_{\mu\nu\mu'} &=& \frac{1}{2} \frac{\partial}{\partial x^\mu} \square g_{\mu'\nu}
   - \frac{1}{2} \frac{\partial^2}{\partial x^\mu \partial x^{\mu'}}
   \frac{\partial g_{\nu\rho}}{\partial x_\rho} \nonumber \\
   && \hspace{-3em} - \, \frac{1}{2} \Gamma^\sigma_{\mu'\mu} \bigg(
   \frac{1}{\tilde{g}} \square g_{\sigma\nu}
   + \frac{\partial}{\partial x^\nu } \frac{\partial g_{\sigma\rho}}{\partial x_\rho}
   - \frac{\partial}{\partial x^\sigma } \frac{\partial g_{\nu\rho}}{\partial x_\rho}
   \bigg) \nonumber \\
   && \hspace{-3em} + \, \frac{\eta^{\rho\rho'}}{2} \Gamma^\sigma_{\rho\nu} \bigg(
   \frac{\partial^2 g_{\sigma\rho'}}{\partial x^\mu \partial x^{\mu'}}
   - \frac{\partial^2 g_{\mu\rho'}}{\partial x^\sigma \partial x^{\mu'}} \nonumber \\
   && \hspace{0.5em} + \,
   2 \frac{\partial^2 g_{\mu\mu'}}{\partial x^\sigma \partial x^{\rho'}}
   - 2 \frac{\partial^2 g_{\sigma\mu'}}{\partial x^\mu \partial x^{\rho'}}
   - \frac{1}{\tilde{g}} \frac{\partial^2 g_{\sigma\mu}}{\partial x^{\mu'} \partial x^{\rho'}}
   \bigg) \nonumber \\
   && \hspace{-3em} + \, \frac{\eta^{\rho\rho'}}{\tilde{g}} \bigg[ 
   \Gamma^\alpha_{\mu'\mu} \bigg( 
   2 \bar{\Gamma}_{\alpha\rho'\beta} + \bar{\Gamma}_{\beta\rho'\alpha}\bigg)
   - \Gamma^\alpha_{\rho'\mu} \bar{\Gamma}_{\alpha\mu'\beta}
   \bigg] \Gamma^\beta_{\rho\nu} \nonumber \\
   && \hspace{8em} - \; \boxed{\mu \leftrightarrow \nu} = 0 .
\label{compactgeq}
\end{eqnarray}
In view of the antisymmetry of $\Xi_{\mu\nu\mu'}$ implied by the last line of the above equation, we can assume $\mu < \nu$ so that Eq.~(\ref{compactgeq}) provides a total of $24$ equations for the ten components of the symmetric matrix $g_{\mu\nu}$. If we wish to determine the time evolution of the metric from the third-order differential equations (\ref{compactgeq}), we need $30$ initial conditions for the matrix elements $g_{\mu\nu}$ and their first and second time derivatives as well as expressions for the third time derivatives.

Closer inspection of the third-order terms in Eq.~(\ref{compactgeq}) reveals that the six equations $\Xi_{0 m n} = 0$ for $m \le n$ provide the derivatives $\partial^3 g_{mn}/\partial t^3$, but that the remaining equations do not contain any information about $\partial^3 g_{0\mu}/\partial t^3$. Therefore, the remaining $18$ equations constitute constraints for the initial conditions, and we are faced with two tasks: (i) find equations for the time evolution of $g_{0\mu}$, and (ii) show that the constraints are satisfied at all times if they hold initially (or count the additional constraints that need to be satisfied otherwise).

It is not at all trivial to find the number of further constraints arising from the dynamic invariance of the constraints contained in Eq.~(\ref{compactgeq}). A controlled handling of constraints is more straightforward in a Hamiltonian setting. As the canonical Hamiltonian formulation has been elaborated only in the weak-field approximation \cite{hco240}, we sketch the generalizations required for the full, nonlinear theory of composite pure gravity in Appendix~\ref{AppHamiltonian}. As a conclusion, we expect (at least) four physical degrees of freedom remaining in the field equations (\ref{compactgeq}) for $g_{\mu\nu}$. Note that the Hamiltonian approach also provides the natural starting point for a generalization to dissipative systems. In particular, this approach allows us to formulate quantum master equations \cite{BreuerPetru,Weiss,hco199,hco221} and to make composite gravity accessible to the robust framework of dissipative quantum field theory \cite{hcoqft,hco243}.

The issue of missing evolution equations is addressed in the subsequent section. As in the weak-field approximation, coordinate conditions characterizing those coordinate systems in which the composite theory of gravity can be applied provide the missing evolution equations.

\section{Coordinate conditions}\label{seccoco3}
As we have assumed an underlying Minkowski space for developing composite gravity, we need to characterize those coordinate systems in which the theory actually holds. These characteristic coordinate conditions should clearly be Lorentz covariant. Furthermore, the coordinate conditions should provide evolution equations for $g_{0\mu}$ because the field equations (\ref{compactgeq}) determine the third-order time derivatives of $g_{mn}$, but not of $g_{0\mu}$. Therefore, the formulation of appropriate coordinate conditions is an important task. The status of coordinate conditions in composite theory is very different from their status in general relativity, where they have no influence on the physical predictions.

The coordinate conditions should be a set of four Lorentz covariant equations. An appealing form is given by
\begin{equation}\label{divcoordcondsg}
   \frac{\partial g_{\mu\rho}}{\partial x_\rho} =
   \frac{\partial \phi}{\partial x^\mu} ,
\end{equation}
where the potential $\phi$ is often assumed to be proportional to the trace of the metric. To eliminate the need of specifying a potential, we can write the second-order integrability conditions
\begin{equation}\label{divcoordcondsint}
   \frac{\partial}{\partial x^\nu} \frac{\partial g_{\mu\rho}}{\partial x_\rho} =
   \frac{\partial}{\partial x^\mu} \frac{\partial g_{\nu\rho}}{\partial x_\rho} .
\end{equation}
After taking the derivatives with respect to $x_\nu$ and summing over $\nu$, we arrive at the four Lorentz covariant coordinate conditions
\begin{equation}\label{coco3K}
   \square \frac{\partial g_{\mu\rho}}{\partial x_\rho} = K \frac{\partial}{\partial x^\mu}
   \frac{\partial^2 g_{\rho\sigma}}{\partial x_\rho\partial x_\sigma} ,
\end{equation}
actually with $K=1$. Note that it is very appealing to use third-order equations as coordinate conditions because we actually need only expressions for the third time derivatives of $g_{0\mu}$ (stronger, first-order conditions are needed for the Hamiltonian formulation; see Appendix~\ref{AppHamiltonian}). For $K=1$, we would obtain such equations for $g_{0m}$, but not for $g_{00}$. This is the reason why we have introduced the factor $K$ in Eq.~(\ref{coco3K}). For any $K \ne 1$, we obtain the desired four evolution equations for $g_{0\mu}$. Formally, we could stick to the first-order conditions (\ref{divcoordcondsg}), but then the potential $\phi$ would be described by the second-order differential equations
\begin{equation}\label{coco3Kphi}
   \square \phi = 
   K \frac{\partial^2 g_{\rho\sigma}}{\partial x_\rho\partial x_\sigma} ,
\end{equation}
where suitable space-time boundary conditions would be required. Note, however, that for $K \ne 1$, Eqs.~(\ref{divcoordcondsg}) and (\ref{coco3Kphi}) imply
\begin{equation}\label{cococons1}
   \frac{\partial^2 g_{\rho\sigma}}{\partial x_\rho\partial x_\sigma} = 0 ,
\end{equation}
whereas Eq.~(\ref{coco3K}) implies the weaker requirement
\begin{equation}\label{cococons3}
   \square \frac{\partial^2 g_{\rho\sigma}}{\partial x_\rho\partial x_\sigma} = 0 .
\end{equation}

The coordinate conditions (\ref{coco3K}) are an essential new ingredient into the composite theory of gravity. Of course, these coordinate conditions take a particularly simple form for $K=0$, which is a possible choice. Alternatively, we could choose $K=\tilde{g}/(1+\tilde{g})$ because we can then express the coordinate conditions as
\begin{equation}\label{cocoGam}
   \frac{\partial^2 \bar{\Gamma}_{\mu\rho\sigma}}{\partial x_\rho\partial x_\sigma} = 0.
\end{equation}
In the following, we leave the particular choice of $K \ne 1$ open.

From a structural point of view, the coordinate conditions (\ref{coco3K}) have the important advantage that they can be implemented in exactly the same way as the gauge conditions in Yang-Mills theories: one can add a term to the Lagrangian that does not lead to any modification of the field equations, provided that the desired (coordinate or gauge) conditions are imposed as constraints. For the coordinate conditions (\ref{coco3K}), the additional contribution to the Lagrangian is given in Appendix~\ref{AppL4cc}.

\section{Static isotropic solution}\label{secisosol}
The study of static isotropic solutions of composite gravity is of great importance because these solutions provide the predictions for the high-precision tests of general relativity (deflection of light by the sun, anomalous precession of the perihelion of Mercury, gravitational redshift of spectral lines from white dwarf stars, travel time delay for radar signals reflecting off other planets) and the properties of black holes. Therefore, we here discuss these solutions in great detail.

We assume that the static isotropic solutions are of the general form,
\begin{equation}\label{isoxg}
   g_{\mu\nu} = \left( \begin{matrix}
   -\beta & 0 \\
   0 & \alpha \, \delta_{mn} + \xi \, \frac{x_m x_n}{r^2}
   \end{matrix} \right) ,
\end{equation}
with inverse
\begin{equation}\label{isoxginv}
   \bar{g}^{\mu\nu} = \left( \begin{matrix}
   -\frac{1}{\beta} & 0 \\
   0 & \frac{\delta_{mn}}{\alpha} - \frac{\xi}{\alpha(\alpha+\xi)} \, \frac{x_m x_n}{r^2}
   \end{matrix} \right) ,
\end{equation}
where $\alpha$, $\beta$ and $\xi$ are functions of the single variable $r=(x_1^2+x_2^2+x_3^2)^{1/2}$. The static isotropic metric (\ref{isoxg}) is given in terms of the three real-valued functions $\alpha$, $\beta$ and $\xi$. In the original work on the composite theory of gravity (see Sec.~V of \cite{hco231}), we had parametrized these three functions in terms of only two functions $A$ and $B$: $\alpha=1$, $\beta=B$, and $\xi=A-1$. This particular parametrization corresponds to standard quasi-Minkowskian coordinates. A problem with these quasi-Minkowskian coordinates is that it is unclear how they can be generalized to full coordinate conditions for general metrics. The more general form (\ref{isoxg}) of the metric is consistent with the coordinate conditions (\ref{coco3K}). In particular, we do not need to introduce a further function for characterizing the components $g_{0m}$. In general relativity, the form (\ref{isoxg}) of the metric (with $g_{0m}=0$) can be achieved by shifting time by a function depending on $r$ (see Sec.~8.1 of \cite{Weinberg}). Nonzero $g_{0m}$ arise by Lorentz transformation of the metric (\ref{isoxg}) so that the form (\ref{isoxg}) belongs to a particularly simple solution of coordinate conditions and field equations.

The field equations (\ref{compactgeq}) provide two third-order ordinary differential equations involving all three functions $\alpha$, $\beta$ and $\xi$. For $K \ne 1$, the coordinate conditions (\ref{coco3K}) lead to another third-order differential equation relating $\alpha$ and $\xi$, which is actually independent of $K$; only for $K=1$, no further condition arises. In the remainder of this section, we solve the three differential equations for our three unknown functions for $K \ne 1$ by various methods.

\subsection{Robertson expansion}
The high-precision tests for theories of gravity depend on the behavior of the static isotropic solutions at large distances. We therefore construct the so-called Robertson expansion in terms of $1/r$. One obtains the following results,
\begin{equation}\label{Robexpansiona}
   \alpha = 1 + \alpha_1 \frac{r_0}{r} + \alpha_3 \frac{r_0^3}{r^3} + \ldots  ,
\end{equation}
\begin{equation}\label{Robexpansionx}
   \xi = \xi_1 \frac{r_0}{r} + \ldots ,
\end{equation}
and
\begin{equation}\label{Robexpansionb}
   \beta  = 1 - 2\frac{r_0}{r}
   + \big[ 2 + (\tilde{g}-1)(\alpha_1+\xi_1) \big] \, \frac{r_0^2}{2 r^2} + \ldots ,
\end{equation}
where all higher terms indicated by $\ldots$\ in these Robertson expansions are uniquely determined by the dimensionless parameters $\alpha_1$, $\alpha_3$, $\xi_1$ and the coupling constant $\tilde{g}$. However, $\alpha_1$ and $\xi_1$ are not independent but rather related by a cubic algebraic equation with a single real solution establishing a one-to-one relation between $\alpha_1$ and $\xi_1$ (see Appendix~\ref{Appcubeq}). The parameter $r_0$ with dimension of length is determined by the mass at the center creating the static isotropic field, as can be shown by reproducing the limit of Newtonian gravity (see, e.g., Sec.~3.4 of \cite{Weinberg}).

An obvious strategy for finding the dimensionless parameters is to make sure that the high-precision predictions of general relativity are reproduced. This is achieved by choosing
\begin{equation}\label{highprecisionmatch}
   \alpha_1 + \xi_1 = 2 , \qquad \alpha_1 = \tilde{g} .
\end{equation}
Imposing a further relation between $\alpha_1$ and $\xi_1$ is subtle as we have already established the cubic relationship between these parameters given explicitly in Eq.~(\ref{cubiceq}). This implies that the first part of Eq.~(\ref{highprecisionmatch}) can be satisfied only for particular values of the coupling constant $\tilde{g}$. By using Eq.~(\ref{highprecisionmatch}) for eliminating $\alpha_1$ and $\xi_1$ from Eq.~(\ref{cubiceq}), we obtain the following equation for $\tilde{g}$,
\begin{equation}\label{grooteq}
   (4 + 4 \tilde{g} - \tilde{g}^2 - 5 \tilde{g}^3) (2 - \tilde{g}) = 0 .
\end{equation}
Two of the roots of this polynomial equation of degree four are real. In addition to the obvious root $2$, implying $\alpha_1=2$ and $\xi_1=0$, one finds the further real-valued root
\begin{displaymath}
   \frac{1}{15} \big[ (1259 + 30 \sqrt{1509})^{1/3} + (1259 - 30 \sqrt{1509})^{1/3} - 1 \big] ,
\end{displaymath}
which is approximately equal to $1.13164$; although closer to unity, this irrational number seems to be less appealing than the integer $2$. Only for these two values of $\tilde{g}$ composite gravity with the coordinate conditions (\ref{coco3K}) for $K \ne 1$ can reproduce the high-precision predictions of general relativity. Note that the parameter $\alpha_3$ in the expansions (\ref{Robexpansiona})--(\ref{Robexpansionb}) remains undetermined as it is the only term among the listed ones that does not affect the high-precision tests of gravity.

\subsection{Short-distance singularity}
We next focus on singular behavior at small distances, which we expect to describe black holes. A glance at the field equations (\ref{compactgeq}) reveals that any fixed multiple of a solution is another solution of the field equations. For the ``equidimensional'' third-order differential equations determining the functions $\alpha$, $\beta$ and $\xi$ of $r$, we assume the following form,
\begin{equation}\label{singularsol}
   \alpha = \frac{c_\alpha}{r^x} , \qquad 
   \beta = \frac{c_\beta}{r^x} , \qquad 
   \xi = \frac{c_\xi}{r^x} , 
\end{equation}
with constants $c_\alpha$, $c_\beta$, $c_\xi$ and an exponent $x$. We further assume that $c_\alpha$, $c_\beta$, and $x$ are different from zero. For $\tilde{g}=2$, we then find that the field equations and coordinate conditions are equivalent to $c_\xi=0$ and $x=1$. For general $\tilde{g}$, one can verify that the values
\begin{equation}\label{singularsolpar}
   x = \frac{2}{\tilde{g}} , \qquad c_\xi=0 ,
\end{equation}
lead to a static isotropic solution of both field equations and coordinate conditions. Of course, this solution is physically unacceptable as a global solution because it does not converge to the Minkowski metric at large distances. It does, however, characterize the asymptotic singular behavior of physical solutions at short distances.

The exponent $x$ given in Eq.~(\ref{singularsolpar}) speaks strongly in favor of choosing $\tilde{g}=2$ (rather than an irrational value). We then obtain a solution decaying according to a $1/r$ power law, the spatial part of which is a multiple of the three-dimensional unit matrix.

\subsection{Numerical solution}
After discussing the static isotropic solutions at large and small distances from the center, we would now like to consider their behavior over the entire range of $r$. In particular, we are interested in the influence of the so far undetermined parameter $\alpha_3$ in Eq.~(\ref{Robexpansiona}) on the behavior of the solutions.

To explore the full solutions, we solve the field equations and coordinate conditions by numerical integration, starting from a large initial distance $r_{\rm i}$ and then proceeding to smaller values of $r$. Assuming $\tilde{g}=2$, the initial conditions at $r_{\rm i}$ are given by the truncated third-order expansions
\begin{equation}\label{Robexpansiona3}
   \alpha = 1 + 2 \frac{r_0}{r} + \alpha_3 \frac{r_0^3}{r^3} ,
\end{equation}
\begin{equation}\label{Robexpansionx3}
   \xi = - 3 \alpha_3 \frac{r_0^3}{r^3} ,
\end{equation}
\begin{equation}\label{Robexpansionb3}
   \beta  = 1 - 2\frac{r_0}{r} + 2 \frac{r_0^2}{r^2} - 2 \frac{r_0^3}{r^3} .
\end{equation}
These expressions do not only provide the values of the coefficient functions at $r_{\rm i}$, but also their first and second derivatives required for solving the third-order differential equations for the functions $\alpha$, $\beta$ and $\xi$ of $r$. The actual numerical solution is performed with an implicit Runge-Kutta scheme of Mathematica.

\begin{figure}
\centerline{\includegraphics[width=8.5 cm]{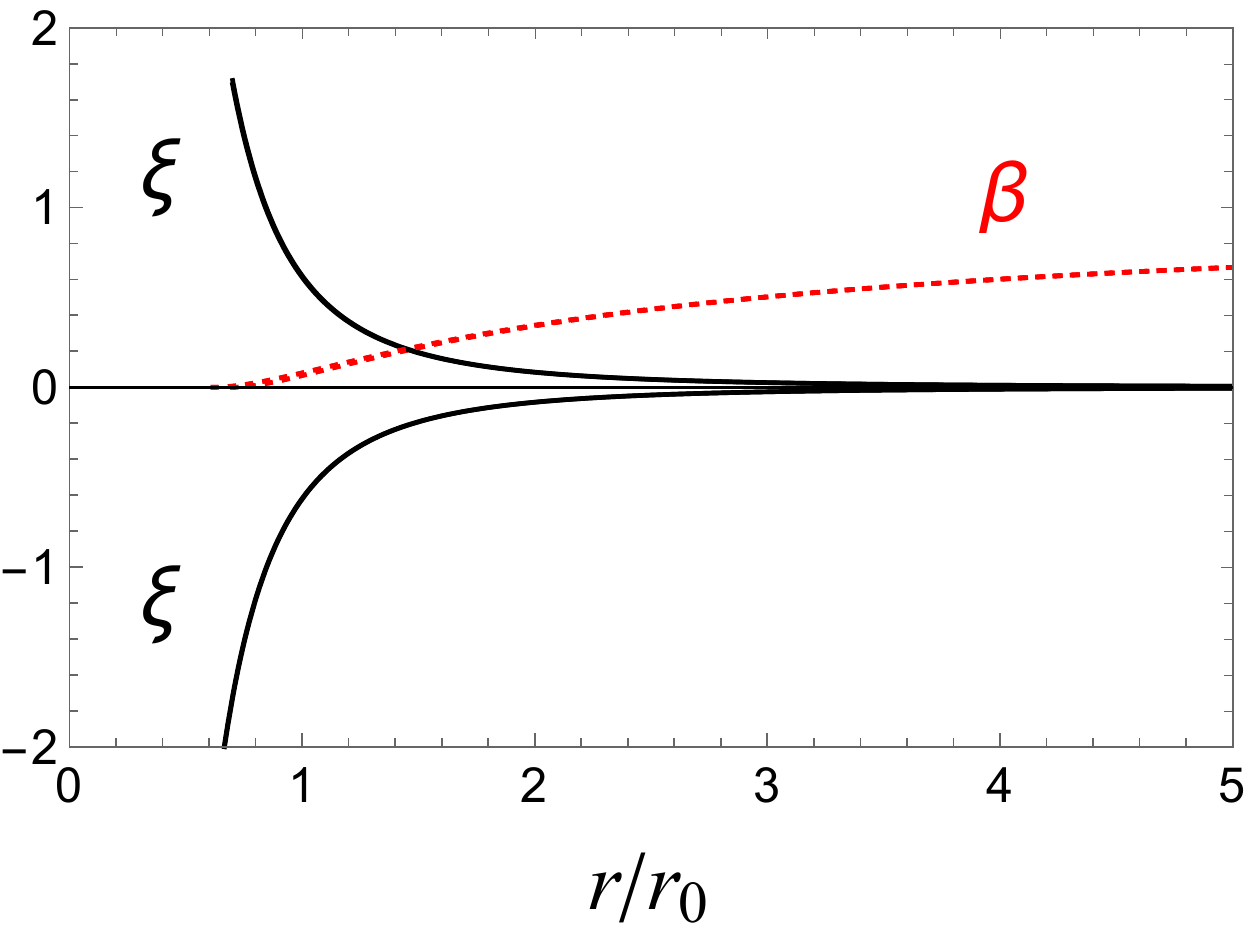}}
\caption[ ]{The functions $\beta$ (dashed line) and $\xi$ (continuous lines) characterizing the temporal and off-diagonal components of the isotropic metric (\ref{isoxg}) obtained from the composite theory for gravity for $\tilde{g}=2$ and $\alpha_3=\pm 0.25$. Positive and negative values of $\xi$ correspond to $\alpha_3=-0.25$ and $\alpha_3=0.25$, respectively.} \label{coordinate_conditions_fig1}
\end{figure}

If $r_{\rm i}$ is sufficiently large, that is, in the range of validity of the asymptotic solutions (\ref{Robexpansiona3})-(\ref{Robexpansionb3}), the numerical solutions are expected to be independent of the choice of $r_{\rm i}$. This expectation is scrutinized in Figure~\ref{coordinate_conditions_fig1}. This figure displays the functions $\beta$ and $\xi$ for the values $\alpha_3=\pm 0.25$ in the conditions (\ref{Robexpansiona3}), (\ref{Robexpansionx3}). The numerical solutions have been calculated for $r_{\rm i}=50$ and $r_{\rm i}=500$, so that each curve for $\xi$ actually consists of two overlapping curves and the anticipated independence of the results of $r_{\rm i}$ is confirmed. The result for $\beta$ actually consists of four curves, which implies that $\xi$ has remarkably little influence on the function $\beta$ until it touches the $r$ axis.

Figure~\ref{coordinate_conditions_fig1} suggests that $\xi$ diverges around the value $r$ at which $\beta$ touches the $r$ axis (and numerical difficulties arise). According to Eqs.~(\ref{singularsol}), (\ref{singularsolpar}), $\xi$ must go to zero for small $r$. The real function $\xi$ might actually end in a cusp singularity and develop a complex branch at smaller $r$ that reaches zero at $r=0$ (see Sec.~V\,C of \cite{hco231}). Alternatively, $\xi$ might jump from $+\infty$ to $-\infty$, or vice versa, to return as a real function to zero at $r=0$, where it started at large $r$ (this kind of behavior is found for the Schwarzschild solution of general relativity; see Sec.~\ref{secSchwarzschild}). To avoid singularities at finite $r$ we from now on assume $\alpha_3=0$, for which $\xi(r)$ is found to be identically zero. Note that singularities would be much more alarming in the composite theory of gravity than in general relativity because they cannot be considered as artifacts (``coordinate singularities'') removable by general coordinate transformations.

\subsection{An exact solution}
\begin{figure}
\centerline{\includegraphics[width=8.5 cm]{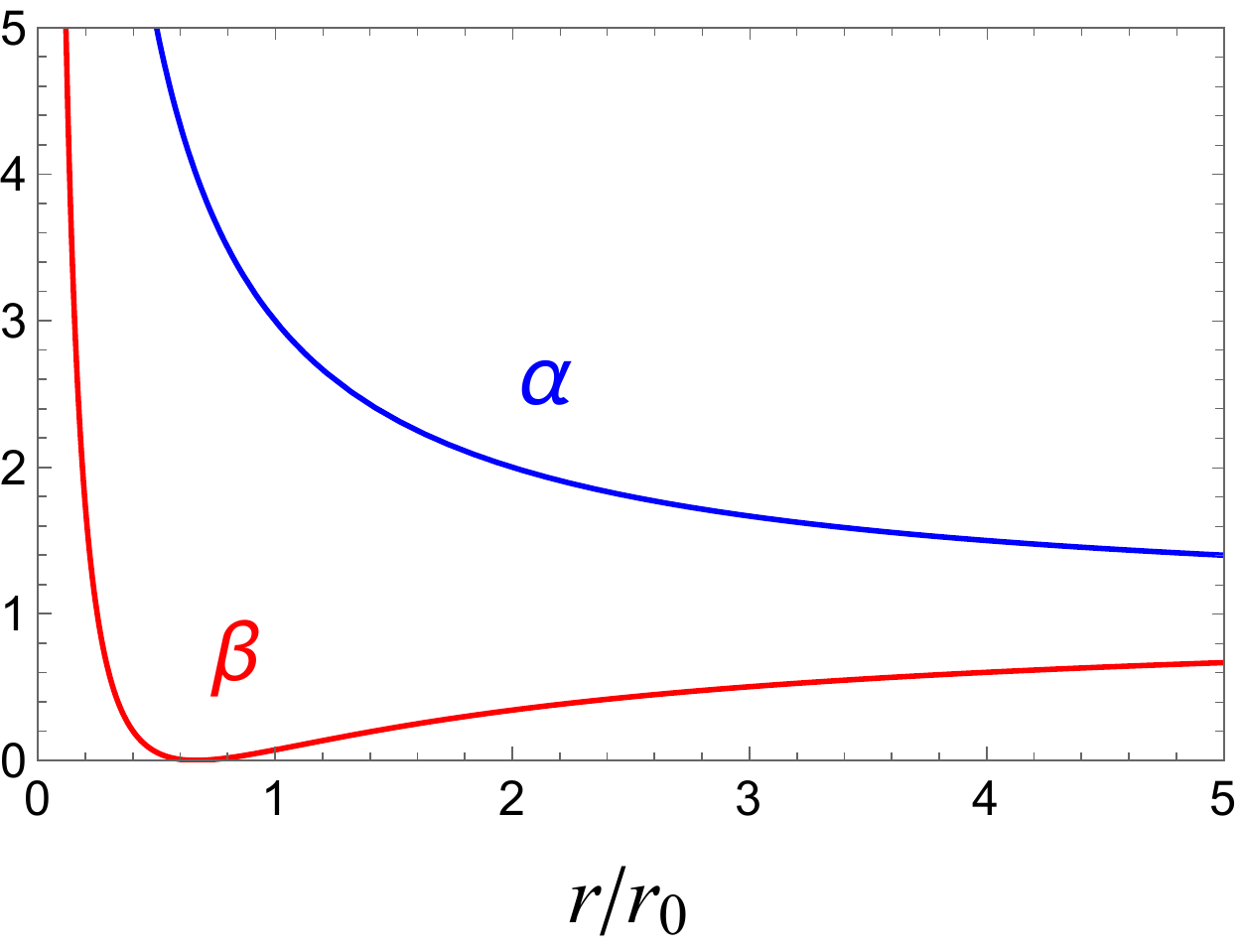}}
\caption[ ]{The exact solutions (\ref{exactisosol}) for the functions $\alpha$ and $\beta$ characterizing the diagonal components of the isotropic metric (\ref{isoxg}) in the composite theory for gravity.} \label{coordinate_conditions_fig2}
\end{figure}

As we have by now fixed the values of the coupling constant ($\tilde{g}=2$) and all the free parameters in the Robertson expansions (\ref{Robexpansiona})-(\ref{Robexpansionb}) ($\alpha_1=2$, $\xi_1=0$, $\alpha_3=0$), there should be a unique static isotropic solution, which is the counterpart of the Schwarzschild solution in general relativity. The Robertson expansions suggest that all higher coefficients $\alpha_n$, $\xi_n$ for $n \ge 2$ vanish, so that $\alpha$ consists of only two terms and $\xi$ vanishes identically, as already noted in the numerical solutions. Then, a closed-form expression for $\beta$ can be found from the field equation
\begin{equation}\label{betaeqex}
   4 r_0^2 \, \beta = r^4 \left( 1 + 2 \frac{r_0}{r} \right) {\beta'}^2 ,
\end{equation}
so that we arrive at the complete solution
\begin{equation}\label{exactisosol}
   \alpha = 1 + 2 \frac{r_0}{r} , \quad \xi = 0 , \quad 
   \beta = \left( 2 - \sqrt{1 + 2 \frac{r_0}{r}} \right)^2 .
\end{equation}
These functions $\alpha$ and $\beta$ are shown in Figure~\ref{coordinate_conditions_fig2}. The present results are qualitatively similar to what was found in previous work on the composite theory of gravity for different coordinate conditions (see Fig.~1 of \cite{hco231}).

Note that $\beta$ is non-negative, vanishes at $r=(2/3) r_0$, and that $\sqrt{\alpha} \pm \sqrt{\beta}=2$, where the $+$ sign holds for $r \ge (2/3) r_0$ and the $-$ sign for $r \le (2/3) r_0$. The only singularities occur at the origin, and they are of the Newtonian $1/r$ type. The most remarkable feature is that $\beta$ reaches a local minimum at $r=(2/3) r_0$, where $\beta$ becomes zero. The observation that the proper time stands still at this distance from the origin is the essence of black-hole behavior in the composite theory of gravity.

An interesting consequence of $\beta=0$ is revealed by considering the curvature scalar
\begin{equation}\label{Rscalardef}
   R = \bar{g}^{\nu\nu'} {R^\mu}_{\nu\mu\nu'} = \tilde{g} \, \bar{g}^{\mu\mu'}
   \bar{g}^{\nu\nu'} \tilde{F}_{\mu\nu \mu'\nu'} = \tilde{g} \, 
   \mbox{$\bar{b}^\mu$}_{\kappa} \mbox{$\bar{b}^\nu$}_{\lambda} {F^{(\kappa\lambda)}}_{\mu\nu} .
\end{equation}
For the isotropic solution given in Eq.~(\ref{exactisosol}), we find
\begin{equation}\label{Rscalardefi}
   R = \frac{16 r_0^2}{r^4 \left( 1 + 2 \frac{r_0}{r} \right)^3
   \left( \sqrt{1 + 2 \frac{r_0}{r}} - 2 \right)} ,
\end{equation}
which implies infinite curvature at $r=(2/3) r_0$ where $\beta$ vanishes, and a change of sign at that point. This is an important insight because, in the weak-field approximation, the curvature scalar and tensor have been explored for the coupling of gravitational field and matter \cite{hco240}. If we want to keep geodesic motion of a mass point in a gravitational field, however, the coupling should be done in terms of a scalar or tensor quantity that is given in terms of  second-derivatives of the metric and vanishes, at least for the static isotropic metric. In this context, the scalar identity (\ref{cococons1}) holding for the static isotropic solution might be useful. A tensorial coupling could be based on the following identity for the static isotropic solution,
\begin{equation}\label{coupltensid}
   \frac{\partial^2 g_{\mu\nu}}{\partial x^\rho \partial x_\rho}
   + \frac{1}{2} \bar{g}^{\rho\rho'} \, \frac{\partial g_{\mu\nu}}{\partial x^\rho} \,
   \frac{\partial g_{\rho'\sigma}}{\partial x_\sigma} = 
   \frac{2 r_0^2}{(r+2r_0)r^3} \, \eta_{\mu\nu} ,
\end{equation}
which implies that the trace-free part of the tensor on the left-hand side vanishes.

\subsection{Comparison to Schwarzschild solution}\label{secSchwarzschild}
The Schwarzschild solution of general relativity in harmonic coordinates is given by (see, e.g., Eq.~(8.2.15) of \cite{Weinberg})
\begin{equation}\label{Schwarzschildiso}
   \alpha = \left( 1 + \frac{r_0}{r} \right)^2 , \quad 
   \xi = \frac{r+r_0}{r-r_0} \, \frac{r_0^2}{r^2} , \quad
   \beta = \frac{r-r_0}{r+r_0} .
\end{equation}
We here compare this solution to the static isotropic solution (\ref{exactisosol}) of composite gravity.

The functions $\alpha$ in Eqs.~(\ref{exactisosol}) and (\ref{Schwarzschildiso}) differ by the term $r_0^2/r^2$. This term does not matter for the high-precision tests. Whereas the singularity at the origin is $1/r$ for composite gravity, it is $1/r^2$ for general relativity. This observation goes nicely with the exponent $x$ in Eqs.~(\ref{singularsol}), (\ref{singularsolpar}), for $\tilde{g}=2$ and $\tilde{g}=1$, respectively, where general relativity corresponds to the torsion-free case $\tilde{g}=1$.

Whereas $\xi$ vanishes in composite gravity, it has a singularity at $r=r_0$ for the Schwarzschild  solution, with a jump from $+\infty$ for $r=r_0^+$ to $-\infty$ for $r=r_0^-$. While this may be considered as a coordinate singularity in general relativity, this would not be possible for a theory in Minkowski space. For the high-precision tests, the absence of a $1/r$ contribution to $\xi$ is crucial.

Also $\beta$ is remarkably different for the two solutions. Whereas $\beta$ is non-negative in composite gravity, it changes sign at $r_0$ for the Schwarzschild solution. Although the two solutions look so different, their truncated third-order expansions (\ref{Robexpansionb3}) coincide. The coincidence of these expansions to order $1/r^2$ is crucial for satisfying the high-precision tests.

\section{Summary and conclusions}\label{secconclusions}
Yang-Mills theories are formulated on a background Minkowski space, and so is the composite theory of gravity that selects a small subset of solutions from the Yang-Mills theory based on the Lorentz group. Such theories are covariant under Lorentz transformations but, unlike general relativity, not under general coordinate transformations. Therefore, it is important to characterize the coordinate systems, in which the composite theory of gravity should be valid, by coordinate conditions. We here propose the Lorentz covariant third-order equations (\ref{coco3K}) for the metric as appealing coordinate conditions that nicely supplement the third-oder differential equations for the composite theory of gravity. Their alternative formulation in Eqs.~(\ref{divcoordcondsg}) and (\ref{coco3Kphi}) shows that we essentially introduce a potential for the divergence of the metric, where the potential itself satisfies a second-order differential equation.

In the original work on composite gravity \cite{hco231}, no general coordinate conditions were given. The static isotropic solution was determined for quasi-Minkowskian coordinates, which are defined only for solutions of this particular type and do not satisfy the new coordinate conditions. Also the coordinate conditions previously used in the complete Hamiltonian formulation of the linearized theory, or weak-field approximation, of composite gravity \cite{hco240} differ from the present proposal. Therefore, previous results are qualitatively similar but quantitatively different from our previous results. The coordinate conditions (\ref{coco3K}) complete the nonlinear theory of pure composite gravity proposed in \cite{hco231}.

The field equations for pure composite gravity can be expressed in a number of different ways. One option is to solve the field equations of the Yang-Mills theory based on the Lorentz group and, in a post-processing step, select those solutions that can be properly expressed in terms of the derivatives of the tetrad variables obtained by decomposing the metric. Alternatively, one can introduce a gauge-invariant connection with torsion and formulate second-order differential equations entirely in terms of those. One is then interested in the solutions for the connection that can be properly expressed in terms of first derivatives of the metric. A final possibility is to write third-order evolution equations directly for the metric.

In the various formulations of the field equations, it is difficult to count the number of degrees of freedom of composite gravity. This difficulty is a consequence of the primary constraints arising from the composition rule of composite theories and serving as a selection principle for the relevant solutions of the underlying Yang-Mills theory. A canonical Hamiltonian formulation on the combined spaces of tetrad and Yang-Mills variables provides the most structured form of both field equations and coordinate conditions. This formulation suggests that composite gravity has four degrees of freedom (whereas the Yang-Mills theory based on the Lorentz group has $24$ degrees of freedom). The Hamiltonian formulation suggests that we deal with two types of constraints: (i) constraints resulting from the composition rule and (ii) gauge constraints. As the former can be handled by Dirac brackets \cite{Dirac50,Dirac58a,Dirac58b} and the latter by the BRST methodology (the acronym derives from the names of the authors of the original papers \cite{BecchiRouetStora76,Tyutin75}; see also \cite{Nemeschanskyetal86,hco229}), the path to quantization of composite gravity is clear. This is a major advantage of an approach starting from the class of Yang-Mills theories, which so successfully describe electro-weak and strong interactions and for which quantization is perfectly understood, and imposing Dirac-type constraints. In addition, this background reveals why composite theories, although they are higher derivative theories, are not prone to Ostrogradsky instabilities.

The fact that just a few degrees of freedom of the Yang-Mills theory based on the Lorentz group survive in the composite theory of gravity is also reflected in its static isotropic solutions. Its Robertson expansion has two free dimensionless parameters in addition to the Yang-Mills coupling constant. For reproducing the high-precison predictios of general relativity, one of the free parameters and the coupling constant ($\tilde{g}=2$) need to be fixed. The remaining dimensionless parameter can be chosen to avoid singularities at finite distances from the origin. A closed-form solution for the static isotropic metric, which plays the same role in composite gravity as the Schwarzschild solution in general relativity, has been found. The solution displays a $1/r$ singularity at the origin but remains finite at all finite values of $r$. The only remarkable feature is $g_{00}=0$ at a particular distance from the origin, which is of the order of the Schwarzschild radius; for all other values of $r$, we have $g_{00}<0$.

This paper develops only the pure theory of gravity. The coupling to matter still needs to be elaborated. For the linearized composite theory of gravity, we had proposed scalar and tensorial coupling mechanisms \cite{hco240}. As the curvature tensor for the static isotropic metric no longer vanishes for the nonlinear theory, which would lead to a deviation from geodesic motion for a coupling based on the curvature tensor, an alternative scalar [see, e.g., Eq.~(\ref{cococons1})] or tensor [see, e.g., Eq.~(\ref{coupltensid})] must be identified for the coupling of the gravitational field to the energy-momentum tensor of matter.

\begin{acknowledgments}
I am grateful for the opportunity to do this work during my sabbatical at the \emph{Collegium Helveticum} in Z\"urich.
\end{acknowledgments}

\appendix

\section{Relation between covariant derivatives}\label{Appcovdevs}
The reformulation of equations for the Yang-Mills theory based on the Lorentz group in the metric language is based on the identity
\begin{equation}\label{supauxf1}
   f_{(\kappa\lambda)}^{bc} B_b C_c =
   \eta^{\kappa'\lambda'} \Big[ B_{(\kappa'\lambda)} C_{(\kappa\lambda')}
   - C_{(\kappa'\lambda)} B_{(\kappa\lambda')} \Big] ,
\end{equation}
which, in view of the definition (\ref{Xtildef}), can be rewritten in the alternative form
\begin{equation}\label{supauxf2}
   {b^\kappa}_\mu {b^\lambda}_\nu f_{(\kappa\lambda)}^{bc} B_b C_c =
   \bar{g}^{\rho\sigma} \Big( \tilde{B}_{\rho\mu} \tilde{C}_{\sigma\nu}
   + \tilde{B}_{\rho\nu} \tilde{C}_{\mu\sigma} \Big) .
\end{equation}
These remarkably simple identities follow from the form of the structure constants of the Lorentz group. After writing the structure constants in the following explicit form (see Table~\ref{tabindexmatch} for the index conventions),
\begin{eqnarray}
   f^{abc} &=& \eta^{\kappa_a \lambda_c} \eta^{\kappa_b \lambda_a} \eta^{\kappa_c \lambda_b}
   - \eta^{\kappa_a \lambda_b} \eta^{\kappa_b \lambda_c} \eta^{\kappa_c \lambda_a}
   \nonumber\\
   &+& \eta^{\kappa_a \kappa_b} \big( \eta^{\kappa_c \lambda_a} \eta^{\lambda_b \lambda_c}
   - \eta^{\kappa_c \lambda_b} \eta^{\lambda_a \lambda_c} \big)
   \nonumber\\
   &+& \eta^{\kappa_a \kappa_c} \big( \eta^{\kappa_b \lambda_c} \eta^{\lambda_a \lambda_b}
   - \eta^{\kappa_b \lambda_a} \eta^{\lambda_b \lambda_c} \big)
   \nonumber\\
   &+& \eta^{\kappa_b \kappa_c} \big( \eta^{\kappa_a \lambda_b} \eta^{\lambda_a \lambda_c}
   - \eta^{\kappa_a \lambda_c} \eta^{\lambda_a \lambda_b} \big) ,
\label{Lorentzstructure}
\end{eqnarray}
the result (\ref{supauxf1}) is obtained by straightforward calculation.

We can now use Eq.~(\ref{supauxf2}) to evaluate the right-hand side of Eq.~(\ref{central}),
\begin{eqnarray}
   {b^\kappa}_\mu {b^\lambda}_\nu \left[
   \frac{\partial X_{(\kappa\lambda)}}{\partial x^{\rho'}}
   + \tilde{g} \, f_{(\kappa\lambda)}^{bc} A_{b\rho'} X_c \right] &=&
   \frac{\partial \tilde{X}_{\mu\nu}}{\partial x^{\rho'}} \nonumber\\
   && \hspace{-12em} + \, \bar{g}^{\rho\sigma} \bigg[ 
   \left( \tilde{g} \tilde{A}_{\rho\mu\rho'} - {b^\kappa}_\rho \,
   \frac{\partial b_{\kappa\mu}}{\partial x^{\rho'}} \right) \tilde{X}_{\sigma\nu}
   \nonumber\\ && \hspace{-9em} + \, 
   \left( \tilde{g} \tilde{A}_{\rho\nu\rho'} - {b^\kappa}_\rho \,
   \frac{\partial b_{\kappa\nu}}{\partial x^{\rho'}} \right) \tilde{X}_{\mu\sigma}
   \bigg] . \qquad 
\label{appcentral}
\end{eqnarray}
By using the composition rule (\ref{nonlinearcompositioncu}) we recover the fundamental relationship (\ref{central}) with the definition (\ref{Gammatildef}) of the connection following from
\begin{equation}\label{Gammatildefx}
   \bar{\Gamma}_{\mu\rho\nu} =
   {b^\kappa}_\mu \, \frac{\partial b_{\kappa\nu}}{\partial x^\rho}
   - \tilde{g} \tilde{A}_{\mu\nu\rho} .
\end{equation}

\section{Alternative expression for field tensor}\label{Appfieldtensor}
From the definitions (\ref{Fdefinition}) and (\ref{Xtildef}) and the fundamental relations (\ref{central}) and (\ref{supauxf2}), we obtain
\begin{eqnarray}
   \tilde{F}_{\mu\nu \mu'\nu'} &=& \frac{\partial \tilde{A}_{\mu\nu\nu'}}{\partial x^{\mu'}}
   - \Gamma^\sigma_{\mu'\mu} \tilde{A}_{\sigma\nu\nu'}
   + \Gamma^\sigma_{\mu'\nu} \tilde{A}_{\sigma\mu\nu'} \nonumber \\
   &-& \frac{\partial \tilde{A}_{\mu\nu\mu'}}{\partial x^{\nu'}}
   + \Gamma^\sigma_{\nu'\mu} \tilde{A}_{\sigma\nu\mu'}
   - \Gamma^\sigma_{\nu'\nu} \tilde{A}_{\sigma\mu\mu'} \nonumber \\
   &-& \tilde{g} \, \bar{g}^{\rho\sigma}
   \Big( \tilde{A}_{\rho\mu\mu'} \tilde{A}_{\sigma\nu\nu'}
   - \tilde{A}_{\rho\nu\mu'} \tilde{A}_{\sigma\mu\nu'} \Big) . \qquad 
\label{Fdefinitiontil}
\end{eqnarray}
By means of Eq.~(\ref{Gammatildefx}), we obtain
\begin{eqnarray}
   \tilde{g} \left( \frac{\partial \tilde{A}_{\mu\nu\nu'}}{\partial x^{\mu'}}
   - \frac{\partial \tilde{A}_{\mu\nu\mu'}}{\partial x^{\nu'}} \right) &=&
   \frac{\partial \bar{\Gamma}_{\mu \mu' \nu}}{\partial x^{\nu'}}
   - \frac{\partial \bar{\Gamma}_{\mu \nu' \nu}}{\partial x^{\mu'}} \nonumber\\
   && \hspace{-6em} + \, \frac{\partial {b^\kappa}_\mu}{\partial x^{\mu'}}
   \frac{\partial b_{\kappa\nu}}{\partial x^{\nu'}}
   - \frac{\partial {b^\kappa}_\mu}{\partial x^{\nu'}}
   \frac{\partial b_{\kappa\nu}}{\partial x^{\mu'}} ,
\label{Gammatildefxder}
\end{eqnarray}
and, again Eq.~(\ref{Gammatildefx}), gives
\begin{equation}\label{bderbder}
   \frac{\partial {b^\kappa}_\mu}{\partial x^{\mu'}}
   \frac{\partial b_{\kappa\nu}}{\partial x^{\nu'}} = \bar{g}^{\rho\sigma}
   ( \bar{\Gamma}_{\rho\mu'\mu} + \tilde{g} \tilde{A}_{\rho\mu\mu'} )
   ( \bar{\Gamma}_{\sigma\nu'\nu} + \tilde{g} \tilde{A}_{\sigma\nu\nu'} ) .
\end{equation}
By combining Eqs.~(\ref{Fdefinitiontil})--(\ref{bderbder}), we finally arrive at
\begin{equation}\label{FdefinitiontilGam}
   \tilde{F}_{\mu\nu \mu'\nu'} = \frac{1}{\tilde{g}} \bigg( 
   \frac{\partial \bar{\Gamma}_{\mu \mu' \nu}}{\partial x^{\nu'}}
   - \frac{\partial \bar{\Gamma}_{\mu \nu' \nu}}{\partial x^{\mu'}}
   + \bar{\Gamma}_{\sigma \mu' \mu} \Gamma^\sigma_{\nu' \nu}
   - \bar{\Gamma}_{\sigma \nu' \mu} \Gamma^\sigma_{\mu' \nu} \bigg) .
\end{equation}
This expression for the field tensor coincides with the one given in Eq.~(\ref{Fdefinitiontilz}) when the definition (\ref{Gammatildef}) of the connection is used.

\section{Field equation for connection}\label{Appfieldeqcon}
By inserting the expression (\ref{R4def}) for the Riemann curvature tensor in terms of the connection, the field equation (\ref{YMfieldeqsR}) for the composite theory of gravity can be written as a second-order differential equation for the connection,
\begin{eqnarray}
   \frac{\partial^2 \Gamma^\mu_{\mu'\nu}}{\partial x_\rho \partial x^\rho}
   &-& \frac{\partial^2 \Gamma^\mu_{\rho\nu}}{\partial x_\rho \partial x^{\mu'}}
   +  \eta^{\rho\rho'} \bigg[
   \Gamma^\sigma_{\mu'\nu} \frac{\partial \Gamma^\mu_{\rho\sigma}}{\partial x^{\rho'}}
   - \Gamma^\mu_{\mu'\sigma} \frac{\partial \Gamma^\sigma_{\rho\nu}}{\partial x^{\rho'}}
   \nonumber \\
   && \hspace{-4em} + \, \Gamma^\mu_{\rho\sigma}
   \bigg( 2 \frac{\partial \Gamma^\sigma_{\mu'\nu}}{\partial x^{\rho'}}
   - \frac{\partial \Gamma^\sigma_{\rho'\nu}}{\partial x^{\mu'}} \bigg)
   - \Gamma^\sigma_{\rho\nu}
   \bigg( 2 \frac{\partial \Gamma^\mu_{\mu'\sigma}}{\partial x^{\rho'}}
   - \frac{\partial \Gamma^\mu_{\rho'\sigma}}{\partial x^{\mu'}} \bigg) \quad\;\;
   \nonumber \\ && \hspace{-4em} + \, 
   \Gamma^\mu_{\rho'\sigma} \Gamma^\sigma_{\rho\sigma'} \Gamma^{\sigma'}_{\mu'\nu}
   + \Gamma^\mu_{\mu'\sigma} \Gamma^\sigma_{\rho\sigma'} \Gamma^{\sigma'}_{\rho'\nu}
   - 2 \Gamma^\mu_{\rho\sigma} \Gamma^\sigma_{\mu'\sigma'} \Gamma^{\sigma'}_{\rho'\nu}
   \bigg] = 0 .
   \nonumber \\ && 
\label{compactGameq}
\end{eqnarray}
Note that $\eta^{\rho\rho'}$ occurs rather than $\bar{g}^{\rho\rho'}$, so that there is no need to know the metric for solving this equation.

\section{Modified Lagrangian}\label{AppL4cc}
The Lagrangian for a pure Yang-Mills theory, including a covariant but gauge breaking term for removing degeneracies associated with gauge invariance (the particular form corresponds to the convenient Feynman gauge), is given by
\begin{equation}\label{LpureYM}
   L = - \int \left(  \frac{1}{4} F^a_{\mu\nu} F_a^{\mu\nu}
   + \frac{1}{2} \frac{\partial A^a_\mu}{\partial x_\mu}
   \frac{\partial A_a^\nu}{\partial x^\nu} \right) d^3x .
\end{equation}
We propose to add the further term
\begin{equation}\label{LpureYMcc}
   L_{\rm cc} = \frac{1}{2} \int \left(
   \frac{\partial^2 g_{\mu\nu}}{\partial x_\mu\partial x^\sigma}
   \frac{\partial^2 {g_\rho}^\nu}{\partial x_\rho\partial x_\sigma}
   - K \frac{\partial^2 g_{\mu\nu}}{\partial x_\mu\partial x_\nu}
   \frac{\partial^2 g_{\rho\sigma}}{\partial x_\rho\partial x_\sigma}
   \right) d^3x ,
\end{equation}
implying the functional derivative
\begin{equation}\label{LpureYMccder}
   \frac{\delta L_{\rm cc}}{\delta g_{\mu\nu}} = \frac{\partial}{\partial x_\mu} \left( 
   \square \frac{\partial {g_\rho}^\nu}{\partial x_\rho} - K \frac{\partial}{\partial x_\nu}
   \frac{\partial^2 g_{\rho\sigma}}{\partial x_\rho\partial x_\sigma} \right) ,
\end{equation}
which vanishes upon imposing the coordinate conditions (\ref{coco3K}) as constraints. If the gauge conditions and the coordinate conditions are imposed as constraints, the above modifications of the Lagrangian for the pure Yang-Mills theory have no effect on the field equations.

\section{Hamiltonian formulation}\label{AppHamiltonian}
For the weak-field approximation of composite gravity, a canonical Hamiltonian formulation with a detailed analysis of all constraints has been given in \cite{hco240}. We here sketch how that approach can be generalized to a full, nonlinear theory of pure gravity selected from the Yang-Mills theory based on the Lorentz group.

The underlying space of the Hamiltonian formulation consists of the tetrad variables ${b^\kappa}_\mu$ and the gauge vector fields $A_{a \nu}$ associated with the Lorentz group as configurational variables, together with their conjugate momenta ${p_\kappa}^\mu$ and $E^{a \nu}$ (where $E_{a j} = F_{a j0}$ and $E_{a 0} = \partial A_{a \mu}/\partial x_\mu$) \cite{hco237,hco240}. This space consists of $80$ fields, but massive constraints arise from the composition rule and gauge invariance so that, in the end, the composite theory of pure gravity turns out to possess only four degrees of freedom.

The generalization of the Hamiltonian (25)--(27) of \cite{hco240} is obtained by introducing the Hamiltonian for the full, nonlinear version of Yang-Mills theory,
\begin{eqnarray}
   H_{\rm pure} &=& \int \bigg[ \frac{1}{2} E^{a \mu} E_{a \mu}
   + \frac{1}{4} F_{a ij} F^{a ij}
   - E^{a 0} \frac{\partial A_{a j}}{\partial x_j} \nonumber \\
   && \hspace{-3em} - \, E^{a j} \left( \frac{\partial A_{a 0}}{\partial x^j}
   + \tilde{g} f_a^{bc} A_{b j} A_{c 0} \right)
   + \mbox{$\dot{b}^\kappa$}_\mu \, {p_\kappa}^\mu \bigg] d^3 x , \qquad 
\label{Hpure}
\end{eqnarray}
where the functional form of the $16$ time derivatives $\mbox{$\dot{b}^\kappa$}_\mu$ in terms of the configurational variables ${b^\kappa}_\mu$ and $A_{a \nu}$ is obtained from $12$ components of the composition rule (\ref{nonlinearcompositioncu}) and the four coordinate conditions (\ref{divcoordcondsg}) (the potential $\phi$ is assumed to be a functional of $g_{\mu\nu}$). For pure gravity without external sources, we can impose the $16$ constraints ${p_\kappa}^\mu = 0$ so that the composite theory consists of selected solutions of the Yang-Mills theory based on the Lorentz group \cite{hco237,hco240}. The terms involving $E^{a 0}$ in the Hamiltonian (\ref{Hpure}) are associated with the gauge breaking term in the Lagrangian (\ref{LpureYM}). Of course, this Hamiltonian implies the canonical evolution equations for the entire set of $80$ fields.

The generalization of the weak-field approximation becomes particularly simple if we introduce the following variables eliminating the nonlinear effects of the coupling constant,
\begin{equation}\label{Abrevedef}
   \breve{A}_{\mu\nu\rho} = \tilde{A}_{\mu\nu\rho}
   - \frac{1}{2 \tilde{g}} \, \Omega_{\mu\nu/\rho} ,
\end{equation}
with
\begin{equation}\label{Omders}
   \Omega_{\mu\nu/\rho} = {b^\kappa}_\mu \, \frac{\partial b_{\kappa\nu}}{\partial x^\rho}
      - \frac{\partial{b^\kappa}_\mu}{\partial x^\rho} \, b_{\kappa\nu} ,
\end{equation}
and
\begin{equation}\label{Ebrevedef0}
   \breve{E}_{\mu\nu 0} = \tilde{E}_{\mu\nu 0}
   - \frac{\eta^{\rho\rho'}}{2\tilde{g}} \Big( 
   \Gamma^\sigma_{\rho\nu} \bar{\Gamma}_{\mu\rho'\sigma}
   - \Gamma^\sigma_{\rho\mu} \bar{\Gamma}_{\nu\rho'\sigma} \Big) ,
\end{equation}
\begin{equation}\label{Ebrevedefj}
   \breve{E}_{\mu\nu j} = \tilde{E}_{\mu\nu j}
   - \frac{1}{\tilde{g}} \bigg(
   \Gamma^\sigma_{j\mu} \bar{\Gamma}_{\sigma 0\nu}
   - \Gamma^\sigma_{j\nu} \bar{\Gamma}_{\sigma 0\mu} \bigg) ,
\end{equation}
as further modifications of the variables $\tilde{A}_{\mu\nu\rho}$ and $\tilde{E}_{\mu\nu\rho}$ defined in Eq.~(\ref{Xtildef}). For example, the composition rule (\ref{nonlinearcompositioncu})
takes the linear form
\begin{equation}\label{compositionql}
   \breve{A}_{\mu\nu\rho} = \frac{1}{2} \left( \frac{\partial g_{\nu\rho}}{\partial x^\mu}
      -\frac{\partial g_{\mu\rho}}{\partial x^\nu} \right) ,
\end{equation}
which corresponds to Eq.~(7) of \cite{hco240} in the symmetric gauge and includes $12$ primary constraints. Also the evolution equations for $\breve{A}_{\mu\nu\rho}$ and hence also the $12$ secondary constraints keep the same form as in the linearized theory (cf.\ Eqs.~(39), (40) and (46), (47) of \cite{hco240}). The $12$ tertiary constraints can be obtained by acting with the operator $\square$ on the primary constraints. The invariance of the tertiary constraints follows from ${p_\kappa}^\mu = 0$. In order to verify the above statements, one needs the identity
\begin{equation}\label{Omsecder}
   \frac{\partial \Omega_{\mu\nu/\rho}}{\partial x_\rho} - \eta^{\rho\rho'}
   \left( \Gamma^\sigma_{\rho\mu} \Omega_{\sigma\nu/\rho'}
   + \Gamma^\sigma_{\rho\nu} \Omega_{\mu\sigma/\rho'} \right) = 0 ,
\end{equation}
which is the counterpart of Eq.~(16) of \cite{hco240} and can be inferred from the gauge invariance of the left-hand side of Eq.~(\ref{Omsecder}). Finally, the $24$ evolution equations for $\breve{E}_{\mu\nu\rho}$ correspond to the field equations given in various forms in Sec.~\ref{secfieldeqs}.

As the structure of the Hamiltonian and the constraints for the full, nonlinear theory is so similar (mostly even formally identical) to the case of the linear weak-field approximation, we expect the same count of $24 + 3 \cdot 12 + 16 =76$ constraints for $2 \cdot (16+24) = 80$ variables. Half of the $24$ constraints associated with gauge invariance result from the gauge conditions $E_{a 0} = \partial A_{a \mu}/\partial x_\mu = 0$, which establish a relationship between the (unphysical) temporal and longitudinal modes of the four-vector potentials. The above arguments suggest that pure composite gravity possesses (at least) four physical degrees of freedom, just as in the thoroughly elaborated special case of the weak-field approximation \cite{hco240}.

\section{A cubic equation}\label{Appcubeq}
The coefficients $\alpha_1$ and $\xi_1$ in the Robertson expansions (\ref{Robexpansiona}), (\ref{Robexpansionx}) are related by the following cubic equation,
\begin{eqnarray}
   10 \xi_1^3
   &+& 10 \tilde{g} \xi_1 (4 \alpha_1^2 + 5 \alpha_1 \xi_1 + 2 \xi_1^2) \nonumber\\
   &-& 5 \tilde{g}^2 \big[ 4 \xi_1 - (\alpha_1 + \xi_1)
   (8 \alpha_1^2 + 9 \alpha_1 \xi_1 - \xi_1^2) \big] \nonumber\\
   &-& 5 \tilde{g}^3 \big[ 4 + 3 (\alpha_1 + \xi_1)^2 \big] (3 \alpha_1 + 2 \xi_1) \nonumber\\
   &+& \tilde{g}^4 (\alpha_1 + \xi_1) (36 + 11 (\alpha_1 + \xi_1)^2) = 0 .
\label{cubiceq}
\end{eqnarray}
Its only real solution for $\alpha_1$ in terms of $\xi_1$ is given by
\begin{eqnarray}
   \alpha_1 &=& \Big[ \Big(w_3 + \sqrt{w_3^2 - w_2^3}\Big)^{1/3}
   + w_2 \Big(w_3 + \sqrt{w_3^2 - w_2^3}\Big)^{-1/3} \nonumber\\
   && \hspace{-2em} - \, \xi_1 (40 + 85 \tilde{g} - 120 \tilde{g}^2 + 33 \tilde{g}^3) \Big] /
   \big[ 3 \tilde{g} (40 - 45 \tilde{g} + 11 \tilde{g}^2) \big] ,
   \nonumber \\ &&
\label{appcubicalpha1}
\end{eqnarray}
with
\begin{eqnarray}
   w_2 &=& 36 \tilde{g}^3 \big( 200 - 345 \tilde{g} + 190 \tilde{g}^2 - 33 \tilde{g}^3 \big)
   \nonumber\\ &+& 5 \big(
   320 + 160 \tilde{g} - 85 \tilde{g}^2 - 282 \tilde{g}^3 + 111 \tilde{g}^4 \big) \xi_1^2 ,
   \qquad 
\label{appcubicw2}
\end{eqnarray}
and
\begin{eqnarray}
   w_3 &=& -5 \xi_1 \Big[ 108 \tilde{g}^4 \big( 520 - 985 \tilde{g} + 633 \tilde{g}^2
   - 155 \tilde{g}^3 + 11 \tilde{g}^4) \nonumber\\
   &+& \big( 12800 + 52800 \tilde{g} - 82200 \tilde{g}^2 - 10735 \tilde{g}^3 \nonumber\\
   && + 63045 \tilde{g}^4 - 33273 \tilde{g}^5 + 5427 \tilde{g}^6 \big) \xi_1^2 \Big] .
\label{appcubicw3}
\end{eqnarray}


\begin{thebibliography}{38}%
\makeatletter
\providecommand \@ifxundefined [1]{%
 \@ifx{#1\undefined}
}%
\providecommand \@ifnum [1]{%
 \ifnum #1\expandafter \@firstoftwo
 \else \expandafter \@secondoftwo
 \fi
}%
\providecommand \@ifx [1]{%
 \ifx #1\expandafter \@firstoftwo
 \else \expandafter \@secondoftwo
 \fi
}%
\providecommand \natexlab [1]{#1}%
\providecommand \enquote  [1]{``#1''}%
\providecommand \bibnamefont  [1]{#1}%
\providecommand \bibfnamefont [1]{#1}%
\providecommand \citenamefont [1]{#1}%
\providecommand \href@noop [0]{\@secondoftwo}%
\providecommand \href [0]{\begingroup \@sanitize@url \@href}%
\providecommand \@href[1]{\@@startlink{#1}\@@href}%
\providecommand \@@href[1]{\endgroup#1\@@endlink}%
\providecommand \@sanitize@url [0]{\catcode `\\12\catcode `\$12\catcode
  `\&12\catcode `\#12\catcode `\^12\catcode `\_12\catcode `\%12\relax}%
\providecommand \@@startlink[1]{}%
\providecommand \@@endlink[0]{}%
\providecommand \url  [0]{\begingroup\@sanitize@url \@url }%
\providecommand \@url [1]{\endgroup\@href {#1}{\urlprefix }}%
\providecommand \urlprefix  [0]{URL }%
\providecommand \Eprint [0]{\href }%
\providecommand \doibase [0]{http://dx.doi.org/}%
\providecommand \selectlanguage [0]{\@gobble}%
\providecommand \bibinfo  [0]{\@secondoftwo}%
\providecommand \bibfield  [0]{\@secondoftwo}%
\providecommand \translation [1]{[#1]}%
\providecommand \BibitemOpen [0]{}%
\providecommand \bibitemStop [0]{}%
\providecommand \bibitemNoStop [0]{.\EOS\space}%
\providecommand \EOS [0]{\spacefactor3000\relax}%
\providecommand \BibitemShut  [1]{\csname bibitem#1\endcsname}%
\let\auto@bib@innerbib\@empty
\bibitem [{\citenamefont {Yang}\ and\ \citenamefont
  {Mills}(1954)}]{YangMills54}%
  \BibitemOpen
  \bibfield  {author} {\bibinfo {author} {\bibfnamefont {C.~N.}\ \bibnamefont
  {Yang}}\ and\ \bibinfo {author} {\bibfnamefont {R.~L.}\ \bibnamefont
  {Mills}},\ }\bibfield  {title} {\enquote {\bibinfo {title} {Conservation of
  isotopic spin and isotopic gauge invariance},}\ }\href {\doibase
  10.1103/PhysRev.96.191} {\bibfield  {journal} {\bibinfo  {journal} {Phys.\
  Rev.}\ }\textbf {\bibinfo {volume} {96}},\ \bibinfo {pages} {191--195}
  (\bibinfo {year} {1954})}\BibitemShut {NoStop}%
\bibitem [{\citenamefont {Utiyama}(1956)}]{Utiyama56}%
  \BibitemOpen
  \bibfield  {author} {\bibinfo {author} {\bibfnamefont {R.}~\bibnamefont
  {Utiyama}},\ }\bibfield  {title} {\enquote {\bibinfo {title} {Invariant
  theoretical interpretation of interaction},}\ }\href {\doibase
  10.1103/PhysRev.101.1597} {\bibfield  {journal} {\bibinfo  {journal} {Phys.\
  Rev.}\ }\textbf {\bibinfo {volume} {101}},\ \bibinfo {pages} {1597--1607}
  (\bibinfo {year} {1956})}\BibitemShut {NoStop}%
\bibitem [{\citenamefont {Yang}(1974)}]{Yang74}%
  \BibitemOpen
  \bibfield  {author} {\bibinfo {author} {\bibfnamefont {C.~N.}\ \bibnamefont
  {Yang}},\ }\bibfield  {title} {\enquote {\bibinfo {title} {Integral formalism
  for gauge fields},}\ }\href {\doibase 10.1103/PhysRevLett.33.445} {\bibfield
  {journal} {\bibinfo  {journal} {Phys.\ Rev.\ Lett.}\ }\textbf {\bibinfo
  {volume} {33}},\ \bibinfo {pages} {445--447} (\bibinfo {year}
  {1974})}\BibitemShut {NoStop}%
\bibitem [{\citenamefont {Blagojevi{\'c}}\ and\ \citenamefont
  {Hehl}(2013)}]{BlagojevicHehl}%
  \BibitemOpen
  \bibinfo {editor} {\bibfnamefont {M.}~\bibnamefont {Blagojevi{\'c}}}\ and\
  \bibinfo {editor} {\bibfnamefont {F.~W.}\ \bibnamefont {Hehl}},\ eds.,\
  \href@noop {} {\emph {\bibinfo {title} {Gauge Theories of Gravitation: A
  Reader with Commentaries}}}\ (\bibinfo  {publisher} {Imperial College
  Press},\ \bibinfo {address} {London},\ \bibinfo {year} {2013})\BibitemShut
  {NoStop}%
\bibitem [{\citenamefont {Capozziello}\ and\ \citenamefont {{De
  Laurentis}}(2011)}]{CapozzielloDeLau11}%
  \BibitemOpen
  \bibfield  {author} {\bibinfo {author} {\bibfnamefont {S.}~\bibnamefont
  {Capozziello}}\ and\ \bibinfo {author} {\bibfnamefont {M.}~\bibnamefont {{De
  Laurentis}}},\ }\bibfield  {title} {\enquote {\bibinfo {title} {Extended
  theories of gravity},}\ }\href {\doibase 10.1016/j.physrep.2011.09.003}
  {\bibfield  {journal} {\bibinfo  {journal} {Phys.\ Rep.}\ }\textbf {\bibinfo
  {volume} {509}},\ \bibinfo {pages} {167--321} (\bibinfo {year}
  {2011})}\BibitemShut {NoStop}%
\bibitem [{\citenamefont {Ivanenko}\ and\ \citenamefont
  {Sardanashvily}(1983)}]{IvanenkoSar83}%
  \BibitemOpen
  \bibfield  {author} {\bibinfo {author} {\bibfnamefont {D.}~\bibnamefont
  {Ivanenko}}\ and\ \bibinfo {author} {\bibfnamefont {G.}~\bibnamefont
  {Sardanashvily}},\ }\bibfield  {title} {\enquote {\bibinfo {title} {The gauge
  treatment of gravity},}\ }\href {\doibase 10.1016/0370-1573(83)90046-7}
  {\bibfield  {journal} {\bibinfo  {journal} {Phys.\ Rep.}\ }\textbf {\bibinfo
  {volume} {94}},\ \bibinfo {pages} {1--45} (\bibinfo {year}
  {1983})}\BibitemShut {NoStop}%
\bibitem [{\citenamefont {{\"O}ttinger}(2018{\natexlab{a}})}]{hco235}%
  \BibitemOpen
  \bibfield  {author} {\bibinfo {author} {\bibfnamefont {H.~C.}\ \bibnamefont
  {{\"O}ttinger}},\ }\bibfield  {title} {\enquote {\bibinfo {title}
  {{H}amiltonian formulation of a class of constrained fourth-order
  differential equations in the {O}strogradsky framework},}\ }\href {\doibase
  10.1088/2399-6528/aaf6f2} {\bibfield  {journal} {\bibinfo  {journal}
  {J.~Phys.\ Commun.}\ }\textbf {\bibinfo {volume} {2}},\ \bibinfo {pages}
  {125006} (\bibinfo {year} {2018}{\natexlab{a}})}\BibitemShut {NoStop}%
\bibitem [{\citenamefont {{\"O}ttinger}(2019)}]{hco237}%
  \BibitemOpen
  \bibfield  {author} {\bibinfo {author} {\bibfnamefont {H.~C.}\ \bibnamefont
  {{\"O}ttinger}},\ }\bibfield  {title} {\enquote {\bibinfo {title} {Natural
  {H}amiltonian formulation of composite higher derivative theories},}\ }\href
  {\doibase 10.1088/2399-6528/ab3634} {\bibfield  {journal} {\bibinfo
  {journal} {J.~Phys.\ Commun.}\ }\textbf {\bibinfo {volume} {3}},\ \bibinfo
  {pages} {085001} (\bibinfo {year} {2019})}\BibitemShut {NoStop}%
\bibitem [{\citenamefont {{\"O}ttinger}(2020{\natexlab{a}})}]{hco231}%
  \BibitemOpen
  \bibfield  {author} {\bibinfo {author} {\bibfnamefont {H.~C.}\ \bibnamefont
  {{\"O}ttinger}},\ }\bibfield  {title} {\enquote {\bibinfo {title} {Composite
  higher derivative theory of gravity},}\ }\href {\doibase
  10.1103/PhysRevResearch.2.013190} {\bibfield  {journal} {\bibinfo  {journal}
  {Phys.\ Rev.\ Research}\ }\textbf {\bibinfo {volume} {2}},\ \bibinfo {pages}
  {013190} (\bibinfo {year} {2020}{\natexlab{a}})}\BibitemShut {NoStop}%
\bibitem [{\citenamefont {Giovanelli}(2020)}]{Giovanelli20}%
  \BibitemOpen
  \bibfield  {author} {\bibinfo {author} {\bibfnamefont {M.}~\bibnamefont
  {Giovanelli}},\ }\bibfield  {title} {\enquote {\bibinfo {title} {Nothing but
  coincidences. {T}he point-coincidence argument and {E}instein’s struggle
  with the meaning of coordinates in physics},}\ }\href
  {http://philsci-archive.pitt.edu/16830/1/COINCIDENCES FINAL.pdf} {\bibfield
  {journal} {\bibinfo  {journal} {Euro.\ Jnl.\ Phil.\ Sci.}\ }\textbf {\bibinfo
  {volume} {10}},\ \bibinfo {pages} {under review} (\bibinfo {year}
  {2020})}\BibitemShut {NoStop}%
\bibitem [{\citenamefont {{\"O}ttinger}(2020{\natexlab{b}})}]{hco240}%
  \BibitemOpen
  \bibfield  {author} {\bibinfo {author} {\bibfnamefont {H.~C.}\ \bibnamefont
  {{\"O}ttinger}},\ }\bibfield  {title} {\enquote {\bibinfo {title}
  {Mathematical structure and physical content of composite gravity in
  weak-field approximation},}\ }\href {\doibase 10.1103/PhysRevD.102.064024}
  {\bibfield  {journal} {\bibinfo  {journal} {Phys.\ Rev.\ D}\ }\textbf
  {\bibinfo {volume} {102}},\ \bibinfo {pages} {064024} (\bibinfo {year}
  {2020}{\natexlab{b}})}\BibitemShut {NoStop}%
\bibitem [{\citenamefont {Peskin}\ and\ \citenamefont
  {Schroeder}(1995)}]{PeskinSchroeder}%
  \BibitemOpen
  \bibfield  {author} {\bibinfo {author} {\bibfnamefont {M.~E.}\ \bibnamefont
  {Peskin}}\ and\ \bibinfo {author} {\bibfnamefont {D.~V.}\ \bibnamefont
  {Schroeder}},\ }\href@noop {} {\emph {\bibinfo {title} {An Introduction to
  Quantum Field Theory}}}\ (\bibinfo  {publisher} {Perseus Books},\ \bibinfo
  {address} {Reading, MA},\ \bibinfo {year} {1995})\BibitemShut {NoStop}%
\bibitem [{\citenamefont {Weinberg}(2005)}]{WeinbergQFT2}%
  \BibitemOpen
  \bibfield  {author} {\bibinfo {author} {\bibfnamefont {S.}~\bibnamefont
  {Weinberg}},\ }\href {\doibase 10.1017/CBO9781139644174} {\emph {\bibinfo
  {title} {Modern Applications}}},\ \bibinfo {series} {The Quantum Theory of
  Fields}, Vol.~\bibinfo {volume} {2}\ (\bibinfo  {publisher} {Cambridge
  University Press},\ \bibinfo {address} {Cambridge},\ \bibinfo {year}
  {2005})\BibitemShut {NoStop}%
\bibitem [{\citenamefont {{\"O}ttinger}(2018{\natexlab{b}})}]{hco229}%
  \BibitemOpen
  \bibfield  {author} {\bibinfo {author} {\bibfnamefont {H.~C.}\ \bibnamefont
  {{\"O}ttinger}},\ }\bibfield  {title} {\enquote {\bibinfo {title} {{BRST}
  quantization of {Y}ang-{M}ills theory: {A} purely {H}amiltonian approach on
  {F}ock space},}\ }\href {\doibase 10.1103/PhysRevD.97.074006} {\bibfield
  {journal} {\bibinfo  {journal} {Phys.\ Rev.\ D}\ }\textbf {\bibinfo {volume}
  {97}},\ \bibinfo {pages} {074006} (\bibinfo {year}
  {2018}{\natexlab{b}})}\BibitemShut {NoStop}%
\bibitem [{\citenamefont {Jim{\'e}nez}\ \emph {et~al.}(2019)\citenamefont
  {Jim{\'e}nez}, \citenamefont {Heisenberg},\ and\ \citenamefont
  {Koivisto}}]{Jimenezetal19}%
  \BibitemOpen
  \bibfield  {author} {\bibinfo {author} {\bibfnamefont {J.~B.}\ \bibnamefont
  {Jim{\'e}nez}}, \bibinfo {author} {\bibfnamefont {L.}~\bibnamefont
  {Heisenberg}}, \ and\ \bibinfo {author} {\bibfnamefont {T.~S.}\ \bibnamefont
  {Koivisto}},\ }\bibfield  {title} {\enquote {\bibinfo {title} {The
  geometrical trinity of gravity},}\ }\href {\doibase 10.3390/universe5070173}
  {\bibfield  {journal} {\bibinfo  {journal} {Universe}\ }\textbf {\bibinfo
  {volume} {5}},\ \bibinfo {pages} {173} (\bibinfo {year} {2019})}\BibitemShut
  {NoStop}%
\bibitem [{\citenamefont {Weinberg}(1972)}]{Weinberg}%
  \BibitemOpen
  \bibfield  {author} {\bibinfo {author} {\bibfnamefont {S.}~\bibnamefont
  {Weinberg}},\ }\href@noop {} {\emph {\bibinfo {title} {Gravitation and
  Cosmology, Principles and Applications of the General Theory of
  Relativity}}}\ (\bibinfo  {publisher} {Wiley},\ \bibinfo {address} {New
  York},\ \bibinfo {year} {1972})\BibitemShut {NoStop}%
\bibitem [{\citenamefont {Ostrogradsky}(1850)}]{Ostrogradsky1850}%
  \BibitemOpen
  \bibfield  {author} {\bibinfo {author} {\bibfnamefont {M.}~\bibnamefont
  {Ostrogradsky}},\ }\bibfield  {title} {\enquote {\bibinfo {title}
  {M{\'e}moires sur les {\'e}quations diff{\'e}rentielles, relatives au
  probl{\`e}me des isop{\'e}rim{\`e}tres},}\ }\href@noop {} {\bibfield
  {journal} {\bibinfo  {journal} {Mem.\ Acad.\ St.~Petersbourg}\ }\textbf
  {\bibinfo {volume} {6}},\ \bibinfo {pages} {385--517} (\bibinfo {year}
  {1850})}\BibitemShut {NoStop}%
\bibitem [{\citenamefont {Woodard}(2015)}]{Woodard15}%
  \BibitemOpen
  \bibfield  {author} {\bibinfo {author} {\bibfnamefont {R.~P.}\ \bibnamefont
  {Woodard}},\ }\bibfield  {title} {\enquote {\bibinfo {title}
  {{O}strogradsky's theorem on {H}amiltonian instability},}\ }\href {\doibase
  10.4249/scholarpedia.32243} {\bibfield  {journal} {\bibinfo  {journal}
  {Scholarpedia}\ }\textbf {\bibinfo {volume} {10}},\ \bibinfo {pages} {32243}
  (\bibinfo {year} {2015})}\BibitemShut {NoStop}%
\bibitem [{\citenamefont {j.~Chen}\ \emph {et~al.}(2013)\citenamefont
  {j.~Chen}, \citenamefont {Fasiello}, \citenamefont {Lim},\ and\ \citenamefont
  {Tolley}}]{Chenetal13}%
  \BibitemOpen
  \bibfield  {author} {\bibinfo {author} {\bibfnamefont {T.}~\bibnamefont
  {j.~Chen}}, \bibinfo {author} {\bibfnamefont {M.}~\bibnamefont {Fasiello}},
  \bibinfo {author} {\bibfnamefont {E.~A.}\ \bibnamefont {Lim}}, \ and\
  \bibinfo {author} {\bibfnamefont {A.~J.}\ \bibnamefont {Tolley}},\ }\bibfield
   {title} {\enquote {\bibinfo {title} {Higher derivative theories with
  constraints: Exorcising {O}strogradski's ghost},}\ }\href {\doibase
  10.1088/1475-7516/2013/02/042} {\bibfield  {journal} {\bibinfo  {journal}
  {J.~Cosmol.\ Astropart.\ Phys.}\ }\textbf {\bibinfo {volume} {02}},\ \bibinfo
  {pages} {042} (\bibinfo {year} {2013})}\BibitemShut {NoStop}%
\bibitem [{\citenamefont {Raidal}\ and\ \citenamefont
  {Veerm{\"a}e}(2017)}]{RaidalVeermae17}%
  \BibitemOpen
  \bibfield  {author} {\bibinfo {author} {\bibfnamefont {M.}~\bibnamefont
  {Raidal}}\ and\ \bibinfo {author} {\bibfnamefont {H.}~\bibnamefont
  {Veerm{\"a}e}},\ }\bibfield  {title} {\enquote {\bibinfo {title} {On the
  quantisation of complex higher derivative theories and avoiding the
  {O}strogradsky ghost},}\ }\href {\doibase 10.1016/j.nuclphysb.2017.01.024}
  {\bibfield  {journal} {\bibinfo  {journal} {Nucl.\ Phys.\ B}\ }\textbf
  {\bibinfo {volume} {916}},\ \bibinfo {pages} {607--626} (\bibinfo {year}
  {2017})}\BibitemShut {NoStop}%
\bibitem [{\citenamefont {Stelle}(1977)}]{Stelle77}%
  \BibitemOpen
  \bibfield  {author} {\bibinfo {author} {\bibfnamefont {K.~S.}\ \bibnamefont
  {Stelle}},\ }\bibfield  {title} {\enquote {\bibinfo {title} {Renormalization
  of higher-derivative quantum gravity},}\ }\href {\doibase
  10.1103/PhysRevD.16.953} {\bibfield  {journal} {\bibinfo  {journal} {Phys.\
  Rev.\ D}\ }\textbf {\bibinfo {volume} {16}},\ \bibinfo {pages} {953--969}
  (\bibinfo {year} {1977})}\BibitemShut {NoStop}%
\bibitem [{\citenamefont {Stelle}(1978)}]{Stelle78}%
  \BibitemOpen
  \bibfield  {author} {\bibinfo {author} {\bibfnamefont {K.~S.}\ \bibnamefont
  {Stelle}},\ }\bibfield  {title} {\enquote {\bibinfo {title} {Classical
  gravity with higher derivatives},}\ }\href {\doibase 10.1007/BF00760427}
  {\bibfield  {journal} {\bibinfo  {journal} {Gen.\ Relat.\ Gravit.}\ }\textbf
  {\bibinfo {volume} {9}},\ \bibinfo {pages} {353--371} (\bibinfo {year}
  {1978})}\BibitemShut {NoStop}%
\bibitem [{\citenamefont {Krasnikov}(1987)}]{Krasnikov87}%
  \BibitemOpen
  \bibfield  {author} {\bibinfo {author} {\bibfnamefont {N.~V.}\ \bibnamefont
  {Krasnikov}},\ }\bibfield  {title} {\enquote {\bibinfo {title} {Nonlocal
  gauge theories},}\ }\href {\doibase 10.1007/BF01017588} {\bibfield  {journal}
  {\bibinfo  {journal} {Theor.\ Math.\ Phys.}\ }\textbf {\bibinfo {volume}
  {73}},\ \bibinfo {pages} {1184--1190} (\bibinfo {year} {1987})}\BibitemShut
  {NoStop}%
\bibitem [{\citenamefont {Grosse-Knetter}(1994)}]{GrosseKnetter94}%
  \BibitemOpen
  \bibfield  {author} {\bibinfo {author} {\bibfnamefont {C.}~\bibnamefont
  {Grosse-Knetter}},\ }\bibfield  {title} {\enquote {\bibinfo {title}
  {Effective {L}agrangians with higher derivatives and equations of motion},}\
  }\href {\doibase 10.1103/PhysRevD.49.6709} {\bibfield  {journal} {\bibinfo
  {journal} {Phys.\ Rev.\ D}\ }\textbf {\bibinfo {volume} {49}},\ \bibinfo
  {pages} {6709--6719} (\bibinfo {year} {1994})}\BibitemShut {NoStop}%
\bibitem [{\citenamefont {Becker}\ \emph {et~al.}(2017)\citenamefont {Becker},
  \citenamefont {Ripken},\ and\ \citenamefont {Saueressig}}]{Beckeretal17}%
  \BibitemOpen
  \bibfield  {author} {\bibinfo {author} {\bibfnamefont {D.}~\bibnamefont
  {Becker}}, \bibinfo {author} {\bibfnamefont {C.}~\bibnamefont {Ripken}}, \
  and\ \bibinfo {author} {\bibfnamefont {F.}~\bibnamefont {Saueressig}},\
  }\bibfield  {title} {\enquote {\bibinfo {title} {On avoiding {O}strogradski
  instabilities within asymptotic safety},}\ }\href {\doibase
  10.1007/JHEP12(2017)121} {\bibfield  {journal} {\bibinfo  {journal} {J.~High
  Energy Phys.}\ }\textbf {\bibinfo {volume} {12}},\ \bibinfo {pages} {121}
  (\bibinfo {year} {2017})}\BibitemShut {NoStop}%
\bibitem [{\citenamefont {Salvio}(2019)}]{Salvio19}%
  \BibitemOpen
  \bibfield  {author} {\bibinfo {author} {\bibfnamefont {A.}~\bibnamefont
  {Salvio}},\ }\bibfield  {title} {\enquote {\bibinfo {title} {Metastability in
  quadratic gravity},}\ }\href {\doibase 10.1103/PhysRevD.99.103507} {\bibfield
   {journal} {\bibinfo  {journal} {Phys.\ Rev.\ D}\ }\textbf {\bibinfo {volume}
  {99}},\ \bibinfo {pages} {103507} (\bibinfo {year} {2019})}\BibitemShut
  {NoStop}%
\bibitem [{\citenamefont {Breuer}\ and\ \citenamefont
  {Petruccione}(2002)}]{BreuerPetru}%
  \BibitemOpen
  \bibfield  {author} {\bibinfo {author} {\bibfnamefont {H.-P.}\ \bibnamefont
  {Breuer}}\ and\ \bibinfo {author} {\bibfnamefont {F.}~\bibnamefont
  {Petruccione}},\ }\href@noop {} {\emph {\bibinfo {title} {The Theory of Open
  Quantum Systems}}}\ (\bibinfo  {publisher} {Oxford University Press},\
  \bibinfo {address} {Oxford},\ \bibinfo {year} {2002})\BibitemShut {NoStop}%
\bibitem [{\citenamefont {Weiss}(2008)}]{Weiss}%
  \BibitemOpen
  \bibfield  {author} {\bibinfo {author} {\bibfnamefont {U.}~\bibnamefont
  {Weiss}},\ }\href@noop {} {\emph {\bibinfo {title} {Quantum Dissipative
  Systems}}},\ \bibinfo {edition} {3rd}\ ed.,\ Series in Modern Condensed
  Matter Physics, Volume~13\ (\bibinfo  {publisher} {World Scientific},\
  \bibinfo {address} {Singapore},\ \bibinfo {year} {2008})\BibitemShut
  {NoStop}%
\bibitem [{\citenamefont {{\"O}ttinger}(2011)}]{hco199}%
  \BibitemOpen
  \bibfield  {author} {\bibinfo {author} {\bibfnamefont {H.~C.}\ \bibnamefont
  {{\"O}ttinger}},\ }\bibfield  {title} {\enquote {\bibinfo {title} {The
  geometry and thermodynamics of dissipative quantum systems},}\ }\href
  {\doibase 10.1209/0295-5075/94/10006} {\bibfield  {journal} {\bibinfo
  {journal} {Europhys.\ Lett.}\ }\textbf {\bibinfo {volume} {94}},\ \bibinfo
  {pages} {10006} (\bibinfo {year} {2011})}\BibitemShut {NoStop}%
\bibitem [{\citenamefont {Taj}\ and\ \citenamefont
  {{\"O}ttinger}(2015)}]{hco221}%
  \BibitemOpen
  \bibfield  {author} {\bibinfo {author} {\bibfnamefont {D.}~\bibnamefont
  {Taj}}\ and\ \bibinfo {author} {\bibfnamefont {H.~C.}\ \bibnamefont
  {{\"O}ttinger}},\ }\bibfield  {title} {\enquote {\bibinfo {title} {Natural
  approach to quantum dissipation},}\ }\href {\doibase
  10.1103/PhysRevA.92.062128} {\bibfield  {journal} {\bibinfo  {journal}
  {Phys.\ Rev.\ A}\ }\textbf {\bibinfo {volume} {92}},\ \bibinfo {pages}
  {062128} (\bibinfo {year} {2015})}\BibitemShut {NoStop}%
\bibitem [{\citenamefont {{\"O}ttinger}(2017)}]{hcoqft}%
  \BibitemOpen
  \bibfield  {author} {\bibinfo {author} {\bibfnamefont {H.~C.}\ \bibnamefont
  {{\"O}ttinger}},\ }\href {\doibase 10.1017/9781108227667} {\emph {\bibinfo
  {title} {A Philosophical Approach to Quantum Field Theory}}}\ (\bibinfo
  {publisher} {Cambridge University Press},\ \bibinfo {address} {Cambridge},\
  \bibinfo {year} {2017})\BibitemShut {NoStop}%
\bibitem [{\citenamefont {Oldofredi}\ and\ \citenamefont
  {{\"O}ttinger}(2021)}]{hco243}%
  \BibitemOpen
  \bibfield  {author} {\bibinfo {author} {\bibfnamefont {A.}~\bibnamefont
  {Oldofredi}}\ and\ \bibinfo {author} {\bibfnamefont {H.~C.}\ \bibnamefont
  {{\"O}ttinger}},\ }\bibfield  {title} {\enquote {\bibinfo {title} {The
  dissipative approach to quantum field theory: Conceptual foundations and
  ontological implications},}\ }\href {\doibase 10.1007/s13194-020-00330-9}
  {\bibfield  {journal} {\bibinfo  {journal} {Euro.\ Jnl.\ Phil.\ Sci.}\
  }\textbf {\bibinfo {volume} {11}},\ \bibinfo {pages} {18} (\bibinfo {year}
  {2021})}\BibitemShut {NoStop}%
\bibitem [{\citenamefont {Dirac}(1950)}]{Dirac50}%
  \BibitemOpen
  \bibfield  {author} {\bibinfo {author} {\bibfnamefont {P.~A.~M.}\
  \bibnamefont {Dirac}},\ }\bibfield  {title} {\enquote {\bibinfo {title}
  {Generalized {H}amiltonian dynamics},}\ }\href {\doibase
  10.4153/CJM-1950-012-1} {\bibfield  {journal} {\bibinfo  {journal} {Canad.\
  J.~Math.}\ }\textbf {\bibinfo {volume} {2}},\ \bibinfo {pages} {129--148}
  (\bibinfo {year} {1950})}\BibitemShut {NoStop}%
\bibitem [{\citenamefont {Dirac}(1958{\natexlab{a}})}]{Dirac58a}%
  \BibitemOpen
  \bibfield  {author} {\bibinfo {author} {\bibfnamefont {P.~A.~M.}\
  \bibnamefont {Dirac}},\ }\bibfield  {title} {\enquote {\bibinfo {title}
  {Generalized {H}amiltonian dynamics},}\ }\href {\doibase
  10.1098/rspa.1958.0141} {\bibfield  {journal} {\bibinfo  {journal} {Proc.\
  Roy.\ Soc.\ A}\ }\textbf {\bibinfo {volume} {246}},\ \bibinfo {pages}
  {326--332} (\bibinfo {year} {1958}{\natexlab{a}})}\BibitemShut {NoStop}%
\bibitem [{\citenamefont {Dirac}(1958{\natexlab{b}})}]{Dirac58b}%
  \BibitemOpen
  \bibfield  {author} {\bibinfo {author} {\bibfnamefont {P.~A.~M.}\
  \bibnamefont {Dirac}},\ }\bibfield  {title} {\enquote {\bibinfo {title} {The
  theory of gravitation in {H}amiltonian form},}\ }\href {\doibase
  10.1098/rspa.1958.0142} {\bibfield  {journal} {\bibinfo  {journal} {Proc.\
  Roy.\ Soc.\ A}\ }\textbf {\bibinfo {volume} {246}},\ \bibinfo {pages}
  {333--343} (\bibinfo {year} {1958}{\natexlab{b}})}\BibitemShut {NoStop}%
\bibitem [{\citenamefont {Becchi}\ \emph {et~al.}(1976)\citenamefont {Becchi},
  \citenamefont {Rouet},\ and\ \citenamefont {Stora}}]{BecchiRouetStora76}%
  \BibitemOpen
  \bibfield  {author} {\bibinfo {author} {\bibfnamefont {C.}~\bibnamefont
  {Becchi}}, \bibinfo {author} {\bibfnamefont {A.}~\bibnamefont {Rouet}}, \
  and\ \bibinfo {author} {\bibfnamefont {R.}~\bibnamefont {Stora}},\ }\bibfield
   {title} {\enquote {\bibinfo {title} {Renormalization of gauge theories},}\
  }\href {\doibase 10.1016/0003-4916(76)90156-1} {\bibfield  {journal}
  {\bibinfo  {journal} {Ann.\ Phys.\ (N.Y.)}\ }\textbf {\bibinfo {volume}
  {98}},\ \bibinfo {pages} {287--321} (\bibinfo {year} {1976})}\BibitemShut
  {NoStop}%
\bibitem [{\citenamefont {Tyutin}(1975)}]{Tyutin75}%
  \BibitemOpen
  \bibfield  {author} {\bibinfo {author} {\bibfnamefont {I.~V.}\ \bibnamefont
  {Tyutin}},\ }\href {https://arxiv.org/abs/0812.0580} {\enquote {\bibinfo
  {title} {Gauge invariance in field theory and statistical physics in operator
  formalism},}\ } (\bibinfo {year} {1975}),\ \bibinfo {note} {preprint of P. N.
  Lebedev Physical Institute, No. 39, 1975, arXiv:0812.0580}\BibitemShut
  {NoStop}%
\bibitem [{\citenamefont {Nemeschansky}\ \emph {et~al.}(1988)\citenamefont
  {Nemeschansky}, \citenamefont {Preitschopf},\ and\ \citenamefont
  {Weinstein}}]{Nemeschanskyetal86}%
  \BibitemOpen
  \bibfield  {author} {\bibinfo {author} {\bibfnamefont {D.}~\bibnamefont
  {Nemeschansky}}, \bibinfo {author} {\bibfnamefont {C.}~\bibnamefont
  {Preitschopf}}, \ and\ \bibinfo {author} {\bibfnamefont {M.}~\bibnamefont
  {Weinstein}},\ }\bibfield  {title} {\enquote {\bibinfo {title} {A {BRST}
  primer},}\ }\href {\doibase 10.1016/0003-4916(88)90233-3} {\bibfield
  {journal} {\bibinfo  {journal} {Ann.\ Phys.\ (N.Y.)}\ }\textbf {\bibinfo
  {volume} {183}},\ \bibinfo {pages} {226--268} (\bibinfo {year}
  {1988})}\BibitemShut {NoStop}%
\end{thebibliography}

%

\end{document}